\documentclass[a4paper,11pt,reqno]{article}
\usepackage{amssymb}

\usepackage[tbtags]{mathtools}

\allowdisplaybreaks[4]
\numberwithin{equation}{section}

\mathtoolsset{
  showonlyrefs,
  showmanualtags
}

\usepackage[utf8x]{inputenc}
\SetUnicodeOption{mathletters}
\newcommand{\myand}{ \ensuremath{ ∧ } }
\newcommand{\myor}{ \ensuremath{ ∨ } }
\DeclareUnicodeCharacter{8341}{_{h}}           
\DeclareUnicodeCharacter{8342}{_{k}}           
\DeclareUnicodeCharacter{8343}{_{l}}           
\DeclareUnicodeCharacter{8344}{_{m}}           
\DeclareUnicodeCharacter{8345}{_{n}}           
\DeclareUnicodeCharacter{8346}{_{p}}           
\DeclareUnicodeCharacter{8347}{_{s}}           
\DeclareUnicodeCharacter{8348}{_{t}}           
\DeclareUnicodeCharacter{10759}{\myand}        
\DeclareUnicodeCharacter{10760}{\myor}         
\DeclareUnicodeCharacter{ 8872}{\vDash}        
\DeclareUnicodeCharacter{ 8877}{\nvDash}       
\DeclareUnicodeCharacter{ 8876}{\nvdash}       
\DeclareUnicodeCharacter{ 8718}{\blacksquare}  
\DeclareUnicodeCharacter{10088}{\bigl(}        
\DeclareUnicodeCharacter{10089}{\bigr)}        
\DeclareUnicodeCharacter{10090}{\Bigl(}        
\DeclareUnicodeCharacter{10091}{\Bigr)}        
\DeclareUnicodeCharacter{10214}{\bigl[}        
\DeclareUnicodeCharacter{10215}{\bigr]}        
\DeclareUnicodeCharacter{10100}{\bigl\{}       
\DeclareUnicodeCharacter{10101}{\bigr\}}       
\DeclareUnicodeCharacter{10096}{\bigl\langle}  
\DeclareUnicodeCharacter{10097}{\bigr\rangle}  
\DeclareUnicodeCharacter{ 9168}{\bigm|}        
\DeclareUnicodeCharacter{ 8230}{\dotsc}        
\DeclareUnicodeCharacter{ 8943}{\dotsb}        
\DeclareUnicodeCharacter{ 8315}{}              
\DeclareUnicodeCharacter{  185}{^{-1}}         

\usepackage[english]{babel}

\usepackage{aliascnt}

\usepackage{amsthm}

\usepackage[bottom,stable,multiple]{footmisc}

\usepackage[
  unicode,
  hyperindex,
  colorlinks,
  linkcolor=black,
  anchorcolor=black,
  citecolor=black,
  filecolor=black,
  menucolor=black,
  runcolor=black,
  urlcolor=black,
  hyperfootnotes=false
]{hyperref}

\theoremstyle{plain}

\newtheorem{theorem}{Theorem}[section]

\newaliascnt{proposition}{theorem}
\newtheorem{proposition}[proposition]{Proposition}
\aliascntresetthe{proposition}

\newaliascnt{lemma}{theorem}
\newtheorem{lemma}[lemma]{Lemma}
\aliascntresetthe{lemma}

\newaliascnt{corollary}{theorem}
\newtheorem{corollary}[corollary]{Corollary}
\aliascntresetthe{corollary}

\theoremstyle{definition}

\newtheorem{definition}{Definition}

\newtheorem*{notation}{Notation}

\newcommand{\mybeginproof}[1]{\noindent\textbf{Proof of {#1}.}}
\newcommand{\myendproof}{\qed}
\newcommand{\myproofnote}[1]{\noindent\textit{Proof.} {#1}}

\newcommand{\eqindent}{ \ensuremath{\quad} }

\newcommand{\rightproof}{\ensuremath{ (→) \; }}
\newcommand{\leftproof}{\ensuremath{ (←) \; }}

\newcommand{\define}{ \ensuremath{ \stackrel{ \mathrm{def} }{ ⇔ } } }
\newcommand{\defeq}{ \ensuremath{ \stackrel{ \mathrm{def} }{ = } } }

\renewcommand{\labelenumi}{(\arabic{enumi})}

\newcommand{\sectref}[1]{\autoref{#1}}
\newcommand{\itemref}[1]{\ensuremath{\text{\autoref{#1}}}}
\newcommand{\byitem}[1]{\ensuremath{\text{by \itemref{#1}}}}
\newcommand{\thref}[1]{\ensuremath{\text{\autoref{#1}}}}
\newcommand{\byth}[1]{\ensuremath{\text{by \thref{#1}}}}

\newcommand{\axiomref}[1]{\ensuremath{\text{\eqref{#1}}}}
\newcommand{\axiomsref}[2]{\ensuremath{\text{\axiomref{#1} and \axiomref{#2}}}}
\newcommand{\byaxiom}[1]{\ensuremath{\text{by \axiomref{#1}}}}
\newcommand{\byaxioms}[2]{\ensuremath{\text{by \axiomref{#1} and \axiomref{#2}}}}
\newcommand{\ruleref}[1]{\ensuremath{\text{\eqref{#1}}}}
\newcommand{\byrule}[1]{\ensuremath{\text{by \ruleref{#1}}}}

\newenvironment{simplelist}{
  \begin{list}{\labelenumi}{
    \setlength{\partopsep}{0cm}
    \setlength{\itemsep}{0cm}
    \setlength{\parsep}{0cm}
    \setlength{\leftmargin}{\leftmargini}
    \setlength{\rightmargin}{0cm}
    \setlength{\listparindent}{0cm}
    \setlength{\itemindent}{0cm}
    \setlength{\labelwidth}{\leftmargin}
    \usecounter{enumi}
  }
}{
  \end{list}
}

\newcommand{\VS}{ \ensuremath{ \mathrm{V_S} } }
\DeclareMathOperator{\Vs}{V_{Set}}
\newcommand{\VR}{ \ensuremath{ \mathrm{V_R} } }
\DeclareMathOperator{\TS}{T_{Set}}
\DeclareMathOperator{\TR}{T_{Rel}}
\DeclareMathOperator{\Formulas}{Form}
\DeclareMathOperator{\Thm}{Thm}
\DeclareMathOperator{\Pow}{\mathcal{P}}
\newcommand{\BCl}{ \ensuremath{ \mathrm{Cl_S} } }
\newcommand{\BUlt}{ \ensuremath{ \mathrm{Ult_S} } }
\newcommand{\BUltz}{ \ensuremath{ ❨\mathrm{Ult_S²}❩₀ } }
\newcommand{\RCl}{ \ensuremath{ \mathrm{Cl_R} } }
\newcommand{\RUlt}{ \ensuremath{ \mathrm{Ult_R} } }
\newcommand{\Rw}{ \ensuremath{ \mathrm{R_w} } }

\newcommand{\cardinality}[1]{ \ensuremath{ \lvert { #1 } \rvert } }
\newcommand{\Cardinality}[1]{ \ensuremath{ \bigl\lvert { #1 } \bigr\rvert } }
\newcommand{\modalbox}[1]{ \ensuremath{ [ {#1} ] } }
\newcommand{\modaldiamond}[1]{ \ensuremath{ \langle {#1} \rangle } }

\title{A system of relational syllogistic incorporating full Boolean reasoning\footnote{A previous version of this paper has been published as \cite{PublishedVersion}.}}

\author{
  \begin{tabular}{ccc}
    \begin{tabular}{c}
      Nikolay Ivanov \\
      \small{\texttt{\href{mailto:naivanov@gmail.com}{naivanov@gmail.com}}}
    \end{tabular} &
    &
    \begin{tabular}{c}
      Dimiter Vakarelov \\
      \small{\texttt{\href{mailto:dvak@fmi.uni-sofia.bg}{dvak@fmi.uni-sofia.bg}}}
    \end{tabular}
  \end{tabular} \\
  \parbox{\textwidth}{\small
    \begin{center}
      Faculty of Mathematics and Informatics, Sofia University\\
      5 James Bourchier Blvd., 1164 Sofia, Bulgaria
    \end{center}
  }
}

\date{ }

\begin{document}

\maketitle

\begin{abstract}
We present a system of relational syllogistic, based on classical propositional logic, having primitives of the following form:
\begin{center}
  \textbf{Some} $a$ are $R$-related to \textbf{some} $b$;

  \textbf{Some} $a$ are $R$-related to \textbf{all} $b$;

  \textbf{All} $a$ are $R$-related to  \textbf{some} $b$;

  \textbf{All} $a$ are $R$-related to \textbf{all} $b$.
\end{center}
Such primitives formalize sentences from natural language like `\textbf{All} students read \textbf{some} textbooks'. Here $a,b$ denote arbitrary sets (of objects), and $R$ denotes an arbitrary binary relation between objects. The language of the logic contains only variables denoting sets, determining the class of set terms, and variables denoting binary relations between objects, determining the class of relational terms. Both classes of terms are closed under the standard Boolean operations. The set of relational terms is also closed under taking the converse of a relation. The results of the paper are the completeness theorem with respect to the intended semantics and the computational complexity of the satisfiability problem.
\end{abstract}


\section{Introduction}

It is a well-known fact that the syllogistic was the first formal theory of logic introduced in Antiquity by Aristotle. It was presented by Łukasiewicz in \cite{Lukasiewicz} as a quantifier-free extension of propositional logic, having as atoms the expressions $A(a,b)$ (All $a$ are $b$)  and $I(a,b)$ (Some $a$ are $b$) and their negations $E(a,b) \define ¬I(a,b)$ and $O(a,b) \define ¬A(a,b)$, where $a,b$ are set (class) variables interpreted in the natural language by noun phrases like `men', `Greeks', `mortal'. An example of an Aristotelian syllogism taken from \cite{Lukasiewicz} is: ``If all men are mortal and all Greeks are men, then all Greeks are mortal''. The specific axioms for $A$ and $I$ from \cite{Lukasiewicz} are (in a different logical notation) the following: L1. $A(a,a)$, L2. $I(a,a)$, L3. $A(b,c) ∧ A(a,b) → A(a,c)$, L4. $A(b,c) ∧ I(b,a) → I(a,c)$. The only rules are Modus Ponens and substitution of a set variable with another set variable. The standard semantics of this language consists of interpreting set variables by arbitrary non-empty sets, $A(a,b)$ as set-inclusion $a⊆b$, and $I(a,b)$ as the overlap relation between sets: $a∩b ≠ ∅$.

Wedberg introduced in \cite{Wedberg} variations of the Aristotelian syllogistic with the operation of complementation $a'$ on set variables interpreted as the Boolean complement of the variable in a given universe. Wedberg's system with unrestricted interpretation on set variables is based on the following axioms (containing only $A$ and complementation because $I(a,b)$ can be defined by $¬A(a,b')$): W1. $A(a,a'')$, W2. $A(a'',a)$, W3. $A(a,b) ∧ A(b,c) → A(a,c)$, W4. $A(a,b) → A(b',a')$. W5. $A(a,a') → A(a,b)$.

Simple Henkin-style completeness and decidability proofs for Łukasiewicz's, Wedberg's and some other classical syllogistic systems were given by Shepherdson in \cite{Shepherdson}. Shepherdson's completeness proofs are based on the notion of partially ordered set $S$ with an operation of complementation `$'$'  satisfying the following axioms for all $a, b ∈ S$: $a'' = a$, $a ≤ b → b' ≤ a'$, and $a ≤ a' → a ≤ b$. Similar structures are now known as \emph{orthoposets} (see \cite{Moss2}). Shepherdson also mentioned in \cite{Shepherdson} systems containing not only complementation on set terms, but also Boolean intersection.

We call the variations of Aristotelian syllogistic, mentioned above, \textbf{classical syllogistics}. All such logics are based on propositional logic, but weaker systems, which do not contain the propositional connectives or contain only negation, have also been considered in the literature. For instance, Moss in \cite{Moss1,Moss2}, motivated mainly with applications of syllogistics to natural languages, considers various syllogistics of classical type, based on languages containing primitives like $A, I, E, O$ with or without complementation on set variables. The corresponding axiomatic systems are based on a number of inference rules with finite sets of atomic premises.

For a long time classical syllogistic has been considered only in introductory courses on elementary logic. Nowadays, however, syllogistic theories, extended and modified in various ways, find applications in different areas, mainly in natural language theory \cite{Givan,Moss1,Moss2,Moss3,NMI,Pratt,Pratt1,Pratt2,Pratt3,Pratt-Third,PrattMoss,TC}, computer science and artificial intelligence \cite{SetSyll,RK,KPG,Orlowska}, generalized quantifiers \cite{GenQuant}, argumentation theory \cite{Argumentation}, cognitive psychology \cite{Psychology1,Psychology2} and others (the list of references is fairly incomplete). Most of the extended syllogistics generalize the standard syllogistic relations $A(a,b)$, $I(a,b)$, $E(a,b)$ and $O(a,b)$ using in their definitions various non-standard quantifiers arising from natural language. Examples: `At least 5 $a$ are $b$', `Exactly 5 $a$ are $b$', `Most $a$ are $b$', `All except 2 $a$ are not $b$', `Many $a$ are not $b$', `Only a few $a$ are not $b$', `Usually some $a$ are not $b$', etc.

Some of the relations between sets $a$ and $b$ are determined by certain relations between their members, expressible by some verbs or verb phrases in the natural language. Examples: `All students \textbf{read} some textbooks', `Some people \textbf{don't like} any cat', `Some vegetarians \textbf{eat} some fish', `All vegetarians \textbf{don't like} any meat', `At least 5 students \textbf{read} all textbooks', etc. Syllogistics studying such expressions are called by Moss and Pratt-Hartmann \cite{PrattMoss,Pratt2} \textbf{relational syllogistics}.

Aristotelian syllogistic and most of its extensions can be considered as logics which fit the structure of natural language. Their primitives like \emph{All A are B}, \emph{Some A are B}, \emph{Most A are B} etc, can be considered as relations between classes (sets of objects), and in this sense syllogistic theories can be treated as certain special theories of classes. On the other hand such primitives express kinds of quantification studied in the theory of generalized quantifiers \cite{DE,GenQuant}. Combining some features from generalized quantifier theory and syllogistic reasoning, a new trend in logic has been developed in recent years, called \emph{natural logic}, or \emph{logic for natural language} with the aim to study logical formalisms which fit well with the structure of natural language (see, for instance, \cite{P} and \cite{E} for other references).

In this paper we introduce a quite rich system of relational syllogistic combining some semantical ideas from the aforementioned papers on relational syllogistics and some technical ideas from \cite{rts,dlrbts}. The language of the logic is similar to the language of Dynamic Logic and contains both set variables and relational variables from which we construct complex terms. Both classes of terms are closed with respect to all Boolean operations while on relational terms we also have the operation `$⁻¹$' of taking the converse. We have five atomic predicates from which we construct the set of formulas using the propositional connectives: $a≤b$, $∃∃(a,b)[α]$, $∀∃(a,b)[α]$, $∃∀(a,b)[α]$, $∀∀(a,b)[α]$. Here $a,b$ are set terms and $α$ is a relational term. The semantical structures are the same as in Dynamic logic $(W, R, v)$, where $R$ is a mapping from relational variables to the set of binary relations on $W$ and $v$ is a mapping which assigns to each set variable a subset of $W$. The semantics of $∀∃(a,b)[α]$ is the following:
\[
 (W,R,v) ⊨ ∀∃(a,b)[α] \text{ iff } ❨∀x ∈ v(a)❩ ❨∃y ∈ v(b)❩ ❨x R(α) y❩ \,.
\]
The semantics of the remaining atomic formulas is analogous. Linguistically these formulas cover the examples like `All students read some textbooks', taking all combinations of `some' and `all', considering subject wide scope reading. Having the operation $α⁻¹$, we may also express in our language the object wide scope reading (see \cite{Moss3} for more details). By means of the Boolean operations on relational terms we may express ``compound verbs'' like `to read but not to write'. Also by `$⁻¹$' we may express the passive voice of the verbs like `is read'. Similarly by means of Boolean operators on set terms we may express compound nouns. Let us note that the signs $∃∀$ in $∃∀(a,b)[α]$, and similarly in the other primitives, are not quantifiers on set or relational variables, but part of the notation of our primitive sentences. We choose this notation just because it corresponds directly to the semantics of these primitives and in this way helps the reader to catch more easily their meaning.

We present a Hilbert-style axiomatic system for the logic based on the axioms of propositional logic, Modus Ponens and several additional finitary inference rules satisfying some syntactic restrictions. The list of axioms contains the finite list of axiom schemes for Boolean algebra plus a finite list of axiom schemes for the basic predicates. In this sense our logic is a quantifier-free first-order system, based on propositional logic. We will not treat in this paper our primitive relations as generalized quantifiers.

Logics with similar rules, which in a sense imitate quantification, and canonical constructions for corresponding completeness proofs are studied in \cite{rts,dlrbts}. We adopt and modify these canonical techniques. There are, however, new difficulties, which have no analogs in \cite{rts,dlrbts}. That is why we need to combine canonical constructions from \cite{rts,dlrbts} with a modification of a copying construction from \cite{bml1,bml2,goranko}. The formulas of our logic have a translation into Boolean Modal Logic (BML) \cite{bml1,bml2} extended with converse on relational terms. We obtain that the complexity of the satisfiability problem for the logic is the same as the complexity of BML \cite{ls}, i.e.~NExpTime if the language contains an infinite number of relational variables, and ExpTime if only a finite number of relational variables is available.

The present paper is an extended version of the first author's master's thesis \cite{Nikolay} and was inspired by \cite{PrattMoss}, especially by the presentation of \cite{PrattMoss} by Moss as an invited lecture at the Conference ``Advances in Modal Logic 2008'' \cite{Moss}.

The paper is organized as follows.

In \sectref{sect-syntax-and-semantics}, we introduce the language and semantics of our logic.

In \sectref{sect-axioms-and-inference-rules}, we list the axioms and inference rules of our logical system. We use the axioms for the contact relation from \cite{rts} and some additional axioms and inference rules which essentially imitate quantifiers in our quantifier-free language.

In \sectref{sect-completeness}, we prove the completeness of our axiomatic system. The proof uses some ideas from the completeness proofs for modal logics of the contact relation \cite{rts} and BML \cite{bml1,bml2}.

In \sectref{sect-complexity}, we discuss the complexity of the satisfiability problem for the logic under consideration and some of its fragments.

\section{Syntax and semantics}\label{sect-syntax-and-semantics}

\subsection{Language}

The language consists of the following sets of symbols:
\begin{simplelist}
  \item an infinite set \VS of set variables;
  \item the set constants $0$ and $1$;
  \item a non-empty set \VR of relational variables such that $\VR∩\VS=∅$;
  \item relational constants $0_R$ and $1_R$;
  \item functional symbols $∩$, $∪$ and $-$ for the operations meet, join and complement;
  \item functional symbol $⁻¹$;
  \item relational symbols $≤$, $∃∃$, $∀∃$, $∀∀$, $∃∀$;
  \item propositional connectives $∧, ∨, ¬, →, ↔$;
  \item propositional constants $⊥$ and $⊤$;
  \item the symbols `(', `)', `[', `]', `,'.
\end{simplelist}

As the language is uniquely determined by the pair $(\VS,\VR)$, we will also call $(\VS,\VR)$ a language. In the first two sections we will keep the language fixed.

Set terms are built from the set constants and set variables by means of the Boolean connectives $∩$, $∪$ and $-$. If $V ⊆ \VS$, we will denote by $\TS(V)$ the set of all set terms with variables from $V$.

We define the set of relational terms $\TR(X)$ with variables in $X ⊆ \VR$ to be the smallest set such that:
\begin{simplelist}
  \item $X ∪ \{0_R, 1_R\} ⊆ \TR(X)$;
  \item If $α ∈ \TR(X)$ then $❴-α, α⁻¹❵ ⊆ \TR(X)$;
  \item If $\{α, β\} ⊆ \TR(X)$ then $\{α∩β, α∪β\} ⊆ \TR(X)$.
\end{simplelist}

Atomic formulas have one of the forms
\begin{align}
 a≤b && ∃∃(a,b)[α] && ∀∃(a,b)[α] && ∀∀(a,b)[α] && ∃∀(a,b)[α] \,,
\end{align}
where $a$ and $b$ are set terms and $α$ is a relational term. Formulas are built from atomic formulas by means of the propositional connectives. We will abbreviate $(a≤b)∧(b≤a)$ as $a=b$ and its negation as $a≠b$. If $V ⊆ \VS$ and $R ⊆ \VR$, we will denote by $\Formulas(V,R)$ the set of all formulas with set variables from the set $V$ and relational variables from $R$.

\subsection{Semantics}

Let $W$ be a set and let $R \colon \VR \to \Pow(W²)$ and $v \colon \VS \to \Pow(W)$ be two functions\footnote{We denote by $\Pow(X)$ the power set of the set $X$.}. $R$ is a valuation of the relational variables, which maps every relational variable to a relation on $W$. The valuation $v$ of the set variables maps set variables to subsets of $W$. We will call the pair $(W,R)$ a frame and the triple $(W,R,v)$ a model. The set $W$ is called the domain of that frame or model.

We extend the function $R$ to the set of all relational terms by defining $R(0_R) = ∅$ and $R(1_R) = W²$ and interpreting the symbols $∩$, $∪$, $-$ and $⁻¹$ by intersection, union, complement in $W²$ and taking the converse of the relations on $W$. We extend the function $v$ to the set of all set terms analogously.

If $M$ is a model and $φ$ is a formula, we will denote the statement that $φ$ is true in $M$ by $M⊨φ$. We define the truth and falsity of atomic formulas in a model $(W,R,v)$ by the following equivalences:
\begin{align}
  (W,R,v) ⊨ a≤b          & ⇔ v(a) ⊆ v(b)                            \\
  (W,R,v) ⊨ ∃∃(a,b)[α] & ⇔ ❨∃x ∈ v(a)❩ ❨∃y ∈ v(b)❩ ❨(x,y) ∈ R(α)❩ \\
  (W,R,v) ⊨ ∀∃(a,b)[α] & ⇔ ❨∀x ∈ v(a)❩ ❨∃y ∈ v(b)❩ ❨(x,y) ∈ R(α)❩ \\
  (W,R,v) ⊨ ∀∀(a,b)[α] & ⇔ ❨∀x ∈ v(a)❩ ❨∀y ∈ v(b)❩ ❨(x,y) ∈ R(α)❩ \\
  (W,R,v) ⊨ ∃∀(a,b)[α] & ⇔ ❨∃x ∈ v(a)❩ ❨∀y ∈ v(b)❩ ❨(x,y) ∈ R(α)❩ \,.
\end{align}
The definition is extended to the set of all formulas according to the standard meaning of the propositional connectives.

\subsection{Relations with natural language semantics}

Linguistically the relational variables are interpreted as transitive verbs, and the set variables -- as count-nouns. The formulas $a≤b$ and $a∩b≠∅$ mean `Every \emph{a} is a \emph{b}' and `Some \emph{a} is a \emph{b}' respectively. To illustrate the meaning of the symbols $  Q₁Q₂ $, let us interpret $a$ as `man', $b$ as `animal', and $α$ as the verb `to like'. We denote the subject wide scope reading and the object wide scope reading of a sentence \dots by (\dots)$_{sws}$ and (\dots)$_{ows}$ respectively.\footnote{If the two readings are equivalent, we omit the annotation.} Then we have the following meanings:

\begin{tabular}{r@{ \emph{means} }l}
  $∃∃(a,b)[α]$ & Some man likes some animal \\
  $∀∀(a,b)[α]$ & Every man likes every animal \\
  $∀∃(a,b)[α]$ & (Every man likes some animal)$_{sws}$ \\
  $∃∀(a,b)[α]$ & (Some man likes every animal)$_{sws}$.
\end{tabular}

To express the object wide scope reading, we need the symbol $⁻¹$ which converts a verb into passive voice. In our example $α⁻¹$ means `to be liked':

\begin{tabular}{r@{ \emph{means} }l}
  $∀∃(b,a)[α⁻¹]$ & (Some man likes every animal)$_{ows}$ \\
  $∃∀(b,a)[α⁻¹]$ & (Every man likes some animal)$_{ows}$.
\end{tabular}

Boolean connectives in set terms formalize negated nouns and the connectives `and' and `or' between nouns. The presence of Boolean operators in relational terms allows us to formalize natural language sentences, which contain negated verbs, as well as compound predicates, such as `sees and hears' ($see ∩ hear$) and `sees, but is not seen' ($see ∩ (-see⁻¹)$).

\section{Axioms and inference rules}\label{sect-axioms-and-inference-rules}
We will use the following notation: If $A$ is a formula or a term, then $\Vs(A)$ denotes the set of set variables which occur in $A$. Also, $\Vs(A₁,…,Aₙ) = ⋃_{i=1}ⁿ \Vs(A_i)$.

The idea behind the list of axioms is the following. Since $∃∃$ is the contact relation from the modal logics of region-based theories of space \cite{rts}, we use the same set of axioms for it. The truth of each of the other three relations $ Q₁Q₂ $ is linked to the truth of $∃∃$ by the following equivalences:
\begin{align}
  & (W,R,v) ⊨   ∀∃(a,b)[α] \\*
  & \eqindent {} ⇔❨∀p⊆W❩ ❪ v(a)∩p = ∅ ⨈  ❨∃x ∈ p   ❩ ❨∃y ∈ v(b)❩ ❨(x,y) ∈ R(α)❩ ❫ \\
  & (W,R,v) ⊨   ∀∀(a,b)[α] \\*
  & \eqindent {} ⇔❨∀p⊆W❩ ❪ v(b)∩p = ∅ ⨈  ❨∀x ∈ v(a)❩ ❨∃y ∈ p   ❩ ❨(x,y) ∈ R(α)❩ ❫ \\
  & (W,R,v) ⊨ ¬ ∃∀(a,b)[α] \\*
  & \eqindent {} ⇔❨∀p⊆W❩ ❪ v(a)∩p = ∅ ⨈ ¬❨∀x ∈ p   ❩ ❨∀y ∈ v(b)❩ ❨(x,y) ∈ R(α)❩ ❫
\end{align}

These equivalences express the following simple statement. If $φ(x)$ is a property of elements $x$ in some set $W$ and $A⊆W$, then $(∀x∈A)φ(x)$ is equivalent to $(∀X⊆W)❨X∩A≠∅ ⇒ (∃x∈X)φ(x)❩$.

Thus, we expressed the universally quantified property $(∀ x∈A)φ(x)$ by the existentially quantified property $(∃x∈X)φ(x)$ and a quantification over sets. Substituting the appropriate formulas in the place of $φ (x)$, we get the above equivalences.

The left-to-right direction of each of these equivalences is a universal formula. We add it to the set of axioms. These are the axioms \axiomref{eq-axiom-ee-ae}, \axiomref{eq-axiom-ae-aa}, \axiomref{eq-axiom-not-aa-not-ea} in the list below. We call them linking axioms, because they link relation symbols $  Q₁Q₂ $ and $ Q₁'Q₂' $, which differ in the first or second quantifier.

The right-to-left directions of the equivalences are not universal formulas. Since we do not have quantifiers in our language, we cannot write these conditions as axioms. Instead, we imitate them by inference rules with a special variable, corresponding to the quantified variable $p$ in the above equivalences, using a technique from \cite{rts}. These are the rules \ruleref{eq-rule-ee-ae}, \ruleref{eq-rule-ae-aa}, \ruleref{eq-rule-not-aa-not-ea} from the list below. We call them linking rules.

We will also use a rule whose only purpose is to derive all formulas of the form
\[
  a≠0 → ∃∃(a,a)⟦(α₁⁻¹∪-α₁) ∩ (α₂⁻¹∪-α₂) ∩ ⋯ ∩ (αₖ⁻¹∪-αₖ)⟧ \,.
\]
These formulas state that the valuation of any relational term of the form $α⁻¹∪(-α)$ must be reflexive. The fact that they are theorems is proved in \thref{th-symmetry} and is used in \thref{th-rel-ult-extension}.

The set of axioms consists of the following groups of formulas:
\begin{enumerate}
  \item A sound and complete set of axiom schemes for propositional calculus;
  \item A set of axioms for Boolean algebra in terms of the relation $≤$;
  \item Axioms for equality:
    \begin{align}
       Q₁Q₂(a,b)[α] ∧ a=c & →  Q₁Q₂(c,b)[α] \tag{$A^=₁$} \label{eq-axiom-equality-first-argument} \\
       Q₁Q₂(a,b)[α] ∧ b=c & →  Q₁Q₂(a,c)[α] \tag{$A^=₂$} \label{eq-axiom-equality-second-argument}
    \end{align}
  \item Axioms for $∃∃$:
    \begin{align}
          a=0 ∨ b=0 & → ¬∃∃(a,b)[α] \tag{$A0$} \label{eq-axiom-if-both-arguments-are-empty-then-not-ee} \\
      ∃∃(a∪b, c)[α] & ↔ ∃∃(a,c)[α] ∨ ∃∃(b,c)[α] \tag{$A^∪₁$} \label{eq-axiom-union-in-first-argument-of-ee} \\
      ∃∃(a, b∪c)[α] & ↔ ∃∃(a,b)[α] ∨ ∃∃(a,c)[α] \tag{$A^∪₂$} \label{eq-axiom-union-in-second-argument-of-ee}
    \end{align}
  \item Linking axioms:
    \begin{align}
        ∀∃(a,b)[α] & → a∩c = 0 ∨  ∃∃(c,b)[α] \tag{$AL₁$} \label{eq-axiom-ee-ae} \\
        ∀∀(a,b)[α] & → b∩c = 0 ∨  ∀∃(a,c)[α] \tag{$AL₂$} \label{eq-axiom-ae-aa} \\
      ¬ ∃∀(a,b)[α] & → a∩c = 0 ∨ ¬∀∀(c,b)[α] \tag{$AL₃$} \label{eq-axiom-not-aa-not-ea}
    \end{align}
  \item Axioms for $0_R$ and $1_R$:
    \begin{align}
      ¬∃∃(a,b)[0_R] & \label{eq-axiom-0_R} \tag{$A0_R$} \\
       ∀∀(a,b)[1_R] & \label{eq-axiom-1_R} \tag{$A1_R$}
    \end{align}
  \item Axioms for $∩$, $∪$, $-$ and $⁻¹$ in relational terms:
    \begin{align}
      & ∀∀(a,b)[α∩β] ↔  ∀∀(a,b)[α] ∧ ∀∀(a,b)[β] \tag{$A∩$} \label{eq-axiom-meet-conjunction} \\
      & ∃∃(a,b)[α∪β] ↔  ∃∃(a,b)[α] ∨ ∃∃(a,b)[β] \tag{$A∪$} \label{eq-axiom-join-disjunction} \\
      & ∀∀(a,b)[-α]  ↔ ¬∃∃(a,b)[α] \tag{$A-$} \label{eq-axiom-complement} \\
      & ∃∃(a,b)[α⁻¹] ↔  ∃∃(b,a)[α] \tag{$A⁻¹$} \label{eq-axiom-converse}
    \end{align}
\end{enumerate}

Inference rules:
\begin{enumerate}
  \item
  \[
    φ, φ→ψ ⊢ ψ \tag{$MP$} \label{eq-rule-MP}
  \]
  \item Special rules imitating quantifiers: If $p ∈ \VS ∖ \Vs(φ,a,b)$ then
    \begin{align}
      & φ → a ∩ p = 0 ∨  ∃∃(p,b)[α] ⊢ φ →  ∀∃(a,b)[α] \tag{$R1$} \label{eq-rule-ee-ae} \\
      & φ → b ∩ p = 0 ∨  ∀∃(a,p)[α] ⊢ φ →  ∀∀(a,b)[α] \tag{$R2$} \label{eq-rule-ae-aa} \\
      & φ → a ∩ p = 0 ∨ ¬∀∀(p,b)[α] ⊢ φ → ¬∃∀(a,b)[α] \tag{$R3$} \label{eq-rule-not-aa-not-ea} \\
      & a ∩ p = 0 ∨ ∃∃(p,p)[α] ⊢ a = 0 ∨ ∃∃(a,a)⟦α ∩ (β⁻¹∪-β)⟧ \tag{$RS$} \label{eq-rule-symmetry}
    \end{align}
    The variable $p$ is called the special variable of the rule.
\end{enumerate}

The notions of proof and theorem are defined in the standard way. We will denote by $\Thm(\VS, \VR)$ the set of all theorems in the language $(\VS,\VR)$.

\begin{proposition}\label{th-soundness}
  All theorems are true in all models.
\end{proposition}
\begin{proof}
  All axioms are true in all models and the rule of MP preserves truth in each model. Each of the special rules preserves validity in each frame, that is: if the premise is true in all valuations on a given frame, then so is the conclusion.
\end{proof}

To illustrate the proof system, we will show a proof of the formula
\[
  ∃∀(a,b)[α]→∀∃(b,a)[α⁻¹] \,.
\]
Let $p,q ∈ \VS$, $p≠q$ and $\{p,q\} ∩\Vs(a,b)=∅
$.
\begin{align}
  ⊢ & ¬∀∀(p,b)[α] ∨ b∩q=0 ∨ ∀∃(p,q)[α] && \byaxiom{eq-axiom-ae-aa} \\
  ⊢ & ¬∀∀(p,b)[α] ∨ b∩q=0 ∨ p∩a=0 ∨ ∃∃(a,q)[α] && \byaxiom{eq-axiom-ee-ae} \\
  ⊢ & a∩p=0 ∨ ¬∀∀(p,b)[α] ∨ b∩q=0 ∨ ∃∃(q,a)[α⁻¹] && \byaxiom{eq-axiom-converse} \\
  ⊢ & ¬∃∀(a,b)[α] ∨ ∀∃(b,a)[α⁻¹] && \text{by \ruleref{eq-rule-not-aa-not-ea} and \ruleref{eq-rule-ee-ae}}
\end{align}

\section{Completeness}\label{sect-completeness}

\subsection{Plan of the completeness proof}

First we review the definition of theories and the construction of maximal theories from consistent sets of formulas in the presence of special rules of inference, which imitate quantifiers (for details, see \cite{rts}). We do not have bound variables in formulas, but we will think of some of the variables as being bound by universal quantifiers. That is why we define a theory as a set of formulas together with a set of unbound variables. The set of formulas will not be closed under arbitrary applications of the special rules, but only under applications of instances of these rules, in which the special variable is among the universally bound variables.

To build a model of a consistent set of formulas, we first need to extend it into a maximal thery. We require that such theories contain for each formula exactly one of the formula itself or its negation, but we also require an analog of Henkin's condition -- if the theory contains the negation of the conclusion of some instance of a special rule (which is existential), it should also contain a negation of the premise of that rule (for some special variable, which may be thought of as a witness for that existential formula).

Our construction of the canonical model is based on the Stone representation theorem for Boolean algebras. It builds the points in the model as ultrafilters in the Boolean algebra of set terms. This gives us the correct interpretation of the Boolean operators on set terms without further effort. The problem is that we do not obtain automatically the intended interpretation of the Boolean operators on relational terms. We explain how we deal with this problem in \sectref{sect-canonical}, after we introduce the necessary notation.

\subsection{Theories}\label{sect-theories}

\begin{definition}
  Let $Γ⊆\Formulas(\VS,\VR)$ and $φ ∈ \Formulas(\VS,\VR)$. We will write $Γ⊢₀φ$ when there is a proof of $φ$ from $Γ$, which does not use the special rules (that is, a proof using only \ruleref{eq-rule-MP}). $Γ$ is called consistent if $Γ∪\Thm(\VS,\VR) ⊬₀ ⊥$.
\end{definition}

\begin{definition}[Theory]
  Let $Γ⊆\Formulas(\VS,\VR)$ and let $V⊆\VS$. We say that the pair $(V,Γ)$ is a theory in the language $(\VS,\VR)$ when the following conditions hold:
  \begin{simplelist}
    \item $\Thm(\VS,\VR) ⊆ Γ$;
    \item If $φ, φ→ψ ∈ Γ$ then $ψ ∈ Γ$;
    \item Let $P(q)$ be a premise of a linking rule, where $q ∈ \VS$ is the special variable of the rule. Let $C$ be the conclusion of that rule, $q ∈ \VS ∖ ❨V∪\Vs(C)❩$ and $P(q) ∈ Γ$. Then $C ∈ Γ$.
  \end{simplelist}

  We say that the theory $(V,Γ)$ is consistent if $⊥ ∉ Γ$.

  We say that the theory $(V,Γ)$ in the language $(\VS,\VR)$ is a good theory if $\cardinality{V} < \cardinality{\VS}$.

  The theory $(V,Γ)$ is called complete if it is consistent and for each formula $φ$ in its language we have either $φ ∈ Γ$ or $¬φ ∈ Γ$.

  The theory $(V,Γ)$ in the language $(\VS,\VR)$ is called rich if for each linking rule with premise $P(q) ∈ \Formulas(\VS,\VR)$ and conclusion $C$ the following implication holds: $C ∉ Γ ⇒ (∃q∈\VS) ❨P(q) ∉ Γ❩$. (The conclusion $C$ uniquely determines $P(q)$ up to a substitution of $q$ with another set variable.)
\end{definition}

\begin{lemma}\label{th-extension-of-consistent-set-to-consistent-theory}
  For every consistent set of formulas $Γ₀$ there exists a consistent theory $T=(V,Γ)$ with $Γ ⊇ Γ₀$.
\end{lemma}

\begin{proof}
  Let $T = ❨\VS, ❴φ ∈ \Formulas(\VS,\VR) ⏐ Γ₀∪\Thm(\VS,\VR) ⊢₀ φ❵❩$.
\end{proof}

\begin{notation}
  We define a relation $⊆$ between theories in the same language:
  \[
    (V₁, Γ₁) ⊆ (V₂, Γ₂) \define V₁ ⊆ V₂ ⨇ Γ₁ ⊆ Γ₂ \,.
  \]
  We will write $φ ∈ (V,Γ)$ if $φ ∈ Γ$.
\end{notation}

We fix a language $(\VS,\VR)$ and introduce the following notation:
\begin{notation}
If $Γ$ is a set of formulas and $φ$ is a formula,
\[
  Γ+φ \defeq ❴ ψ ∈ \Formulas(\VS,\VR) ⏐ φ→ψ ∈ Γ ❵ \,.
\]

If $T = (V,Γ)$ is a theory and $φ$ is a formula,
\[
  T⊕φ \defeq ❨V∪\Vs(φ), Γ+φ❩ \,.
\]
\end{notation}

\begin{lemma}\label{th-step-extension-of-a-good-theory}
  If $T = (V,Γ)$ is a good theory and $φ$ is a formula then:
  \begin{enumerate}
    \item $T⊕φ$ is a good theory, $T ⊆ T⊕φ$ and $φ ∈ T⊕φ$;
    \item $T⊕φ$ is inconsistent $⇔$ $¬φ ∈ Γ$;
    \item If $P(q)$ and $C$ are the premise and conclusion of a linking rule and the theory $T ⊕ ¬C$ is consistent, then there is a set variable $q ∈ \VS ∖ ❨V∪\Vs(C)❩$, such that $T ⊕ ¬C ⊕ ¬P(q)❩$ is a good consistent theory.
  \end{enumerate}
\end{lemma}

\begin{proof}
Straightforward verification.
%
%
%
%
\end{proof}

\begin{lemma}[Lindenbaum]
  Every good consistent theory $T₀=(V₀,Γ₀)$ in a language $(\VS,\VR)$ with $\cardinality{\VR} ≤ \cardinality{\VS}$ is contained in a complete rich theory $T=(V,Γ)$.
\end{lemma}

\begin{proof}
  Let $T₀=(V₀,Γ₀)$ be a good consistent theory. Let $κ = \cardinality{\VS}$ and let $\Formulas(\VS,\VR) = \{ φ_α | α<κ \}$. We will build a sequence of theories $\{ T_α \}_{α<κ}$ with the following properties:
  \begin{enumerate}
    \item $T_α$ is a good consistent theory;
    \item $¬φ_α ∈ T_α$ or $φ_α ∈ T_{α+1}$;
    \item If $φ_α ∈ T_{α+1}$, $φ_α = ¬C$ and $C$ is the conclusion of a linking rule, then there is a set variable $q$, such that the negated premise of the rule $¬P(q)$ belongs to $T_{α+1}$.
  \end{enumerate}
  Suppose that $T_β$ have been defined for $β<α$. We will define $T_α$. We consider the following cases:
  \begin{enumerate}
    \item $α=β+1$ for some $β$ and $T_β = (V_β, Γ_β)$ has already been defined. We need to consider two possibilities for the theory $T_β⊕φ_β$:
    \begin{enumerate}
      \item $T_β⊕φ_β$ is consistent. We have two cases depending on $φ_β$:
      \begin{enumerate}
        \item $φ_β$ does not have the form of a negated conclusion of a linking rule. In this case we define $T_α = T_β⊕φ_β$.
        \item $φ_β = ¬C$ and $C$ is a conclusion of a linking rule. Let $P(q)$ be the premise of that rule. According to \thref{th-step-extension-of-a-good-theory} there is a set variable $q ∈ \VS ∖ ❨V_β∪\Vs(φ_β)❩$ such that $T_β⊕φ_β⊕¬P(q)$ is a good consistent theory. We choose such a variable $q$ and define $T_α = T_β⊕φ_β⊕¬P(q)$.
      \end{enumerate}
      \item $T_β⊕φ_β$ is inconsistent. Then \thref{th-step-extension-of-a-good-theory} tells us that $¬φ_β ∈ Γ_β$. We define $T_α = T_β$.
    \end{enumerate}
    \item $α = ⋃α$. We define $V_α = ⋃\{ V_β | β<α \}$, $Γ_α = ⋃\{ Γ_β | β<α \}$ and $T_α = (V_α, Γ_α)$.
  \end{enumerate}

  It is easy to verify the three properties of $T_α$ stated above by induction on $α$. We define $V = ⋃\{ V_α | α<κ \}$, $Γ = ⋃\{ Γ_α | α<κ \}$ and $T = (V,Γ)$. By the properties of $T_α$ for $α<κ$ it easily follows that $T$ is a complete rich theory.
\end{proof}

The Lindenbaum lemma is only applicable to good theories. That is why we will also need the following lemma:

\begin{lemma}
  Let $T₀=(V,Γ₀)$ be a consistent theory in a language $(\VS₀,\VR)$ and let $\VS⊇\VS₀$ with $\cardinality{\VS}>\cardinality{\VS₀}$ be an extension of $\VS₀$ with a set $\VS ∖ \VS₀$ of new set variables. Then there is a good consistent theory $T=(\VS₀,Γ)$ in the language $(\VS,\VR)$ such that $Γ₀⊆Γ$.
\end{lemma}

\begin{proof}
  Define
  \[
    Γ = ❴ φ ∈ \Formulas(\VS,\VR) ⏐ (∃ψ ∈ Γ₀) ❨ψ→φ ∈ \Thm(\VS,\VR)❩ ❵ \,.
  \]
  It is straightforward to check that $T=(\VS₀,Γ)$ has the desired properties.

%
\end{proof}

\begin{corollary}\label{th-every-consistent-set-is-contained-in}
  \begin{enumerate}
    \item Every consistent set of formulas is contained in a good consistent theory in an extension of the language with a set of new set variables.
    \item\label{th-every-consistent-set-is-contained-in-a-maximal-theory} Every consistent set of formulas is contained in a complete rich theory in an extension of the language with a set of new set variables.
  \end{enumerate}
\end{corollary}

A complete rich theory is also called a maximal theory.

\subsection{Boolean algebras of classes of terms}\label{sect-boolean-algebras-of-classes-of-terms}

Let $S$ be a maximal theory. We will associate with $S$ some equivalence relations in $\TS(\VS)$ and $\TR(\VR)$ and will show that the equivalence classes form Boolean algebras with respect to some naturally defined operations.

\subsubsection{The Boolean algebra of classes of set terms}

We will associate with $S$ a Boolean algebra of classes of set terms. We define the relations $≼$ and $≈$ on $\TS(\VS)$:
\begin{align}
  a≼b & \define a≤b ∈ S & a≈b & \define (a≼b ⨇ b≼a) \,.
\end{align}
The relation $≈$ is an equivalence relation. We denote by $[a]$ the equivalence class of $a$. We denote by $\BCl$ the set of all equivalence classes. We define a relation $≤$ on $\BCl$: $[a]≤[b] \define a≼b$. We define the operations $∩$, $∪$ and $-$ on $\BCl$:
\begin{align}
[a]∩[b] & \defeq [a∩b] & [a]∪[b] & \defeq [a∪b] &
-[a] & \defeq [-a]  \,.
\end{align}

The relation $≤$ and the operations $∩$, $∪$ and $-$ are well-defined. The six-tuple $❨\BCl,∩,∪ ,-,[0],[1]❩$ is a Boolean algebra.

\subsubsection{The Boolean algebra of classes of relational terms}

We define the relations $≼$ and $≈$ on the set of all relational terms:
\begin{align}
  α≼β & \define ❨∀a,b∈\TS(\VS)❩ ❨∃∃(a,b)[α]→∃∃(a,b)[β] ∈ S❩ \\
  α≈β & \define (α≼β ⨇ β≼α) \,.
\end{align}
The intuition behind this definition is that in every model $(W,R,v)$ of $S$ the following implication must hold for arbitrary relational terms $α$ and $β$: $α≼β ⇒ R(α)⊆R(β)$.

The relation $≈$ is an equivalence relation. We denote by $[α]$ the equivalence class of $α$. We denote by $\RCl$ the set of all equivalence classes. We define a relation $≤$ on $\RCl$: $[α]≤[β] \define α≼β$. We define the operations $∩$, $∪$, $-$ and $⁻¹$ on $\RCl$:
\begin{align}
[α]∩[β] & \defeq [α∩β] & [α]∪[β] & \defeq [α∪β] \\
   -[α] & \defeq [-α]  &   [α]⁻¹ & \defeq [α⁻¹] \,.
\end{align}

\begin{proposition}\label{th-rel-bool-algebra}
  The relation $≤$ and the operations $∩$, $∪$, $-$ and $⁻¹$ on $\RCl$ are well-defined. The six-tuple $(\RCl,∩,∪,-,[0_R],[1_R])$ is a Boolean algebra and for arbitrary relational terms $α$ and $β$ we have the equivalence $α ≼ β ⇔ α∪β ≈ β$.
\end{proposition}

\myproofnote{See \sectref{Appendix-proof-of-th-rel-bool-algebra}.}

\begin{lemma}\label{th-bool-rel}
  $[1] ≤ [0] ⇔ [1_R] ≤ [0_R]$.
\end{lemma}

\begin{proof}
  \rightproof Let $1=0 ∈ S$ and $a,b ∈ \TS(\VS)$. Then $a=0 ∈ S$ and $b=0 ∈ S$. By \axiomref{eq-axiom-if-both-arguments-are-empty-then-not-ee}, $¬∃∃(a,b)[1_R] ∈ S$, and hence
  \[
    ∃∃(a,b)[1_R]→∃∃(a,b)[0_R] ∈ S \,.
  \]

  \leftproof Assume that $[1_R] ≤ [0_R]$. Then
  \[
    ∃∃(1,1)[1_R]→∃∃(1,1)[0_R] ∈ S \,.
  \]
  By \axiomref{eq-axiom-0_R}, $¬∃∃(1,1)[0_R] ∈ S$, and hence $¬∃∃(1,1)[1_R] ∈ S$.\\
  By \axiomref{eq-axiom-1_R}, $∀∀(1,1)[1_R] ∈ S$, hence $∀∀(1,1)[1_R] ∧ ¬∃∃(1,1)[1_R] ∈ S$.\\
  Using \axiomsref{eq-axiom-ee-ae}{eq-axiom-ae-aa}, we conclude that $1=0 ∈ S$.
\end{proof}

\subsubsection{The Boolean algebra of symmetric classes of relational terms}

We define an operation $⁻¹$ on $\Pow(\RCl)$: For each $V⊆\RCl$
\[
  V⁻¹ \defeq ❴ [α]⁻¹ ⏐ [α]∈V ❵ \,.
\]

\begin{lemma}
  \begin{enumerate}
    \item If $α ∈ \TR(\VR)$ then $❨α⁻¹❩⁻¹ ≈ α$.
    \item If $x ∈ \RCl$ then $❨x⁻¹❩⁻¹ = x$. If $V ⊆ \RCl$ then $❨V⁻¹❩⁻¹ = V$.
    \item Let $V ⊆ \RCl$. If $V$ is a filter, then so is $V⁻¹$. If $V$ is an ultrafilter, then so is $V⁻¹$.
  \end{enumerate}
\end{lemma}

We will call $x ∈ \RCl$ symmetric if $x = x⁻¹$. Similarly, we will call $V ⊆ \RCl$ symmetric if $V = V⁻¹$.

\begin{lemma}
  The set of symmetric classes of relational terms is a Boolean subalgebra of $(\RCl,∩,∪,-,[0_R],[1_R])$.
\end{lemma}

\begin{lemma}\label{th-symmetry}
 If $a$ is a set term and $α₁, α₂,… ,αₖ$ are relational terms, then the formula $a=0 ∨ ∃∃(a,a)⟦(α₁⁻¹∪-α₁) ∩ (α₂⁻¹∪-α₂) ∩ ⋯ ∩ (αₖ⁻¹∪-αₖ)⟧$ is a theorem.
\end{lemma}

\begin{proof}
Let $p₁, p₂,… ,pₖ$ be different set variables, which do not occur in $a$. By \axiomref{eq-axiom-1_R}, \axiomref{eq-axiom-ae-aa} and \axiomref{eq-axiom-ee-ae}, we have
\begin{align}
  & ⊢ p₁∩⋯∩pₖ∩a=0 ∨ ∃∃(p₁∩⋯∩pₖ∩a, p₁∩⋯∩pₖ∩a)[1_R] \\
  & ⊢ p₁∩⋯∩pₖ∩a=0 ∨ ∃∃(p₁, p₁)[1_R] \qquad \text{by \itemref{th-some-theorems-ee-gt} in \thref{th-some-theorems}} \\
  & ⊢ p₂∩⋯∩pₖ∩a=0 \\*
  & \eqindent   {} ∨ ∃∃(p₂∩⋯∩pₖ∩a, p₂∩⋯∩pₖ∩a)⟦α₁⁻¹∪-α₁⟧ \qquad \byrule{eq-rule-symmetry}
\end{align}
Similarly we obtain
\begin{align}
  & ⊢ p₂∩⋯∩pₖ∩a=0 ∨ ∃∃(p₂, p₂)⟦α₁⁻¹∪-α₁⟧ \\
  & ⊢ p₃∩⋯∩pₖ∩a=0 \\*
  & \eqindent   {} ∨ ∃∃(p₃∩⋯∩pₖ∩a, p₃∩⋯∩pₖ∩a)⟦(α₁⁻¹∪-α₁) ∩ (α₂⁻¹∪-α₂)⟧
\end{align}
Continuing in the same way, we arrive at
\[
  ⊢ a=0 ∨ ∃∃(a,a)⟦(α₁⁻¹∪-α₁) ∩ (α₂⁻¹∪-α₂) ∩ ⋯ ∩ (αₖ⁻¹∪-αₖ)⟧ \,.
\]
\end{proof}

\subsection{Canonical construction}\label{sect-canonical}

Let $S$ be a maximal theory. We will prove that $S$ has a model.

We denote by $M_∅$ the model $❨∅, \VR×\{∅\}, \VS×\{∅\}❩$.

\begin{lemma}\label{th-the-empty-model-is-a-model-of-S}
  If $S$ is a maximal theory and $1=0 ∈ S$, then $M_{∅} ⊨ S$.
\end{lemma}

\begin{proof}
  As $S$ is a maximal theory, it suffices to prove the equivalence
  \[
    M_{∅} ⊨ φ ⇔ φ ∈ S
  \]
  for atomic formulas $φ$.

  \begin{enumerate}
    \item Clearly, all formulas in the form of $a≤b$ belong to $S$ and are true in $M_{∅}$.

    \item $φ$ is $∃∃(a,b)[α]$. Then $M_{∅} ⊭ φ$. By \axiomref{eq-axiom-if-both-arguments-are-empty-then-not-ee} $φ ∉ S$.

    \item\label{th-the-empty-model-is-a-model-of-S-ae} $φ$ is $∀∃(a,b)[α]$. Then $M_{∅} ⊨ φ$. For the sake of contradiction suppose that $φ ∉ S$. There exists a set variable $p$ for which
    \[
      a∩p=0∨∃∃(p,b)[α] ∉ S \,.
    \]
    This is a contradiction, since $a∩p=0 ∈ S$.

    \item $φ$ is $∀∀(a,b)[α]$ or $∃∀(a,b)[α]$. Follows from the above and \thref{th-dual-quantifiers}.
  \end{enumerate}
\end{proof}

We will now consider the case when $1=0 ∉ S$.

We denote by $\BUlt$ the set of ultrafilters of the Boolean algebra
\[
  ❨\BCl,∩,∪,-,[0],[1]❩ \,.
\]
Similarly, we denote by $\RUlt$ the set of ultrafilters of the Boolean algebra
\[
  (\RCl,∩,∪,-,[0_R],[1_R]) \,.
\]
Since $[1]≰[0]$, the set $\BUlt$ is non-empty. By \thref{th-bool-rel}, we have also $\RUlt≠∅$.

If $Q$ is a quantifier and $F(a)$ is a statement about set terms, such that $a≈b$ implies $F(a)⇔F(b)$, we will use $❨Q[a]∈\BCl❩F(a)$ as an abbreviation for $(Qx∈\BCl)(∃a∈x)F(a)$. We will also use a similar notation for statements about relational terms.

\begin{notation}
  If $a∈\TS(\VS)$, we denote by $[a) = ❴ x∈\BCl ⏐ [a]≤x ❵$ the smallest filter containing $[a]$. Similarly, if $α∈\TR(\VR)$, we denote by $[α) = ❴ x∈\RCl ⏐ [α]≤x ❵$ the smallest filter containing $[α]$.
\end{notation}

We will explain the ideas which lead us to the definition of the canonical model of $S$. We may attempt to define the model as $M₀ = (W₀,R₀,v₀)$, where:
\begin{simplelist}
 \item $W₀ = \BUlt$;
 \item For each relational term $α$ let
 \begin{multline}
  R₀(α) = \Bigl\{ (U₁,U₂) ∈ \BUlt² \Bigm| \Bigr. \\*
          \Bigl.  ❨∀[a₁]∈U₁❩ ❨∀[a₂]∈U₂❩ ❨∃∃(a₁,a₂)[α] ∈ S❩ \Bigr\} \,;
 \end{multline}
 \item For each set variable $p$ let $v₀(p) = \{ x∈W₀ | [p]∈x \}$.
\end{simplelist}
Then, for arbitrary set terms $a, b$ and an arbitrary relational term $α$ we have:
\begin{align}
 & a≤b ∈ S ⇔ v₀(a)⊆v₀(b) \\
 & Q₁Q₂(a,b)[α] ∈ S ⇔ ❨Q₁x∈v(a)❩❨Q₂y∈v(b)❩❨(x,y)∈R₀(α)❩ \,.
\end{align}
There is, however, a problem with this model. The function $R₀$ may not follow the correct semantics of the Boolean operators -- we do not necessarily have $R₀(α∩β) = R₀(α)∩R₀(β)$ and $R₀(α) ∩ R₀(-α) = ∅$. To build the canonical model, we first define a relation $R⁰_V$ on $W₀$ for each relational ultrafilter $V$, such that the valuation of each relational term $α$ in $M₀$ will be a union of such relations. For each $V ⊆ \RCl$ we define a relation $R⁰_V ⊆ {\Pow(\BCl)}²$:
\begin{multline}
  R⁰_V = \Bigl\{ (U₁,U₂) ∈ {\Pow(\BCl)}² \Bigm| \Bigr.                                \\*
          \Bigl. ❨∀[α]∈V❩ ❨∀[a₁]∈U₁❩ ❨∀[a₂]∈U₂❩ ❨∃∃(a₁,a₂)[α] ∈ S❩ \Bigr\} \,.
\end{multline}
Now for each $α ∈ \TR(\VR)$ we have:
\[
  R₀(α) = W₀² ∩ ⋃❴ R⁰_V ⏐ V∈\RUlt ⨇ [α]∈V ❵ \,.
\]
We should replace $R⁰_V$ with another relation $R_V$ defined for each $V∈\RUlt$, such that $R_{V'} ∩ R_{V''} = ∅$ for different $V', V'' ∈ \RUlt$. The universe of our model will consist of a number of copies of $W₀$. If $(x,y) ∈ R⁰_{V'} ∩ R⁰_{V''}$ for different $V', V'' ∈ \RUlt$, we will have some copies $x',x''$ of $x$ and some copies $y',y''$ of $y$, such that $(x',y') ∈ R_{V'}$ and $(x'',y'') ∈ R_{V''}$.

\begin{notation}
  Let $F₁$ and $F₂$ be filters in the Boolean algebra of $\BCl$ and let $G$ be a filter in the Boolean algebra of $\RCl$. We will use the following notation:
  \begin{align}
    I_{F₁,F₂} & = \Bigl\{ [α] ∈ \RCl \Bigm| ❨∃[a₁]∈F₁❩ ❨∃[a₂]∈F₂❩ ❨∃∃(a₁,a₂)[α]∉S❩ \Bigr\} \\
    I_{G,F₂}  & = \Bigl\{ [a₁]∈ \BCl \Bigm| ❨∃[α] ∈G ❩ ❨∃[a₂]∈F₂❩ ❨∃∃(a₁,a₂)[α]∉S❩ \Bigr\} \\
    I_{F₁,G}  & = \Bigl\{ [a₂]∈ \BCl \Bigm| ❨∃[a₁]∈F₁❩ ❨∃[α] ∈G ❩ ❨∃∃(a₁,a₂)[α]∉S❩ \Bigr\}
  \end{align}
\end{notation}

It is easy to check that the $I$'s are ideals in the respective Boolean algebras.

\begin{lemma}\label{th-filter-to-ultrafilter-extension}
  Let $F₁$ and $F₂$ be filters in the Boolean algebra of $\BCl$ and let $G$ be a filter in the Boolean algebra of $\RCl$. If $(F₁,F₂) ∈ R⁰_G$, then there are $U₁,U₂ ∈ \BUlt$ and $V ∈ \RUlt$ such that $F₁⊆U₁$, $F₂⊆U₂$, $G⊆V$ and $(U₁,U₂) ∈ R⁰_V$.
\end{lemma}

\begin{proof}
  We use the equivalences
  \[
    (F₁,F₂)∈R⁰_G ⇔ F₁∩I_{G,F₂}=∅ ⇔ G∩I_{F₁,F₂}=∅ ⇔ F₂∩I_{F₁,G}=∅
  \]
  and apply the separation theorem for filter-ideal pairs in Boolean algebras three times.
\end{proof}

Let us first exclude the symbol $⁻¹$ from the language. To construct the relations $R_V$, we need the following lemma:

\begin{lemma}
  Let $(U₁,U₂) ∈ \BUlt²$. Then:
  \begin{enumerate}
    \item $(U₁,U₂) ∈ R⁰_{[1_R)}$.
    \item There is a $V ∈ \RUlt$ such that $(U₁,U₂) ∈ R⁰_V$.
  \end{enumerate}
\end{lemma}

\begin{proof}
  \begin{enumerate}
    \item Suppose this is not true. Since $S$ is a complete theory,
    \[
      ❨∃[a₁] ∈ U₁❩❨∃[a₂] ∈ U₂❩❨¬∃∃(a₁,a₂)[1_R] ∈ S ❩ \,.
    \]
    By the axiom for $1_R$, $∀∀(a₁,a₂)[1_R] ∈ S$. Using the linking axioms, we derive $a₁=0 ∨ a₂=0 ∈ S$. Since $S$ is a complete theory, $a₁=0 ∈ S$ or $a₂=0 ∈ S$, hence $[a₁]=[0]$ or $[a₂]=[0]$. This is a contradiction, as $U₁$ and $U₂$ are ultrafilters. Thus, $(U₁,U₂) ∈ R⁰_{[1_R)}$.
    \item By the previous item, $(U₁,U₂) ∈ R⁰_{[1_R)}$. By \thref{th-filter-to-ultrafilter-extension}, there is a $V ∈ \RUlt$ such that $(U₁,U₂) ∈ R⁰_V$.
  \end{enumerate}
\end{proof}

For each pair $(U₁,U₂) ∈ \BUlt²$ we choose one $V ∈ \RUlt$, such that $(U₁,U₂)∈R⁰_V$, and denote it by $V_{U₁,U₂}$.

The canonical model $M=(W,R,v)$ corresponding to $S$ is defined as follows: The domain is $W = \BUlt × \RUlt$. If $x∈W$, we denote by $x₁$ and $x₂$ its first and second component respectively. For each $p ∈ \VS$ we define
\[
  v(p) = ❴ x∈ W ⏐ [p] ∈ x₁ ❵ \,.
\]
It is easy to check that for all set terms $a$ we have $v(a) = ❴ x∈W ⏐ [a] ∈ x₁ ❵$. For each $V∈\RUlt$ we define a relation $R_V ⊆ W²$:
\begin{multline}
  R_V = \Bigl\{ (x,y)∈W² \Bigm| ❨(x₁,y₁)∈R⁰_{y₂} ⨇ V=y₂❩ \Bigr.   \\*
        \Bigl. ⨈ ❨ (x₁,y₁)∉R⁰_{y₂} ⨇ V=V_{x₁,y₁} ❩ \Bigr\} \,.
\end{multline}
That is, if there should be a pair $❨(x₁,\_),(y₁,\_)❩$\footnote{The symbol `$\_$' here denotes an arbitrary element of $\RUlt$.} in $R_V$, we put all pairs $❨(x₁,\_),(y₁,V)❩$ there; if $R_V$ should not contain a pair $❨(x₁,\_),(y₁,\_)❩$, we put all pairs $❨(x₁,\_),(y₁,V)❩$ in $R_{V_{x₁,y₁}}$.

This simple construction suffices to prove the completeness of the proof system without \axiomref{eq-axiom-converse} and \ruleref{eq-rule-symmetry} for the language without $⁻¹$.

When we include the symbol $⁻¹$, however, we need something more sophisticated. The problem is that we should ensure that $R_{V₁} ∩ R_{V₂} = ∅$ for $V₁≠V₂$ while at the same time preserving the property stated in the following lemma:

\begin{lemma}\label{th-R_V-symmetry}
  If $V ⊆ \RCl$, then $R⁰_{V⁻¹} = ❨R⁰_V❩⁻¹$.
\end{lemma}

The decision where to put pairs of points should not be made independently for  $(x,y)$ and $(y,x)$. We should have $(x,y)∈ R_V ⇔ (y,x)∈(R_V)⁻¹$. To this end, we will replace the first disjunct in the above definition of $R_V$ with a condition, in which $V$ does not depend solely on $y₂$, but on a symmetric function of $x₂$ and $y₂$. To define such a function, we number the elements of $\RUlt$ with ordinals and define a symmetric binary operation $⊖$ on them. This is \thref{def-operations} below. We have two different definitions of $⊖$ -- for finite $\cardinality{\RUlt}$ and for infinite $\cardinality{\RUlt}$. We will denote by $V_α$ the element of $\RUlt$ numbered with $α$. We will take the second component of each point of $W$ to be the number (ordinal) of a relational ultrafilter rather than the ultrafilter itself. Let $(x,y)∈W²$ and $n = x₂⊖y₂$.

First we consider the case $x₂≠y₂$. If $(x₁,y₁) ∈ R⁰_{Vₙ} ∖ R⁰_{Vₙ⁻¹}$, we put $(x,y)$ in $R_{Vₙ}$ and $(y,x)$ in $R_{Vₙ⁻¹}$. If $(x₁,y₁) ∈ R⁰_{Vₙ⁻¹} ∖ R⁰_{Vₙ}$, we put $(x,y)$ in $R_{Vₙ⁻¹}$ and $(y,x)$ in $R_{Vₙ}$. In the case when $(x₁,y₁) ∈ R⁰_{Vₙ} ∩ R⁰_{Vₙ⁻¹}$, we need to choose one of $(x,y)$ and $(y,x)$ and then put the chosen pair in $R_{Vₙ}$, while the other one should go to $R_{Vₙ⁻¹}$.

If $x₂=y₂$ or $(x₁,y₁) ∉ R⁰_{Vₙ} ∪R⁰_{Vₙ⁻¹}$, we put the pair $(x,y)$ in the relation corresponding to some relational ultrafilter $V_{x₁,y₁}$, which we choose among those $V∈\RUlt$ for which $(x₁,y₁) ∈R⁰_V$. The reason why we treat the case $x₂=y₂$ along with $(x₁,y₁) ∉ R⁰_{Vₙ} ∪ R⁰_{Vₙ⁻¹}$ rather than putting $(x,y)$ in $R_{Vₙ}$, is that we cannot guarantee that $V_{x₂⊖x₂}$ is symmetric. But we can choose $V_{x₁,x₁}$ to be symmetric according to the following proposition:

\begin{proposition}\label{th-rel-ult-extension}
  If $U ∈ \BUlt$, then:
  \begin{enumerate}
    \item $(U,U) ∈ R⁰_V ⇔ ❨∀[α] ∈ V❩ ❨∀[a] ∈ U❩ ❨∃∃(a,a)[α] ∈ S❩$.
    \item There exists a symmetric $V ∈ \RUlt$ such that $(U,U) ∈ R⁰_V$.
  \end{enumerate}
\end{proposition}

\myproofnote{See \sectref{Appendix-proof-of-th-rel-ult-extension}.}

\begin{definition}\label{def-operations}
Let $κ = \Cardinality{\RUlt}$. We consider two cases for $κ$:

\begin{enumerate}
  \item $κ < ω$. Let $\RUlt = ❴V_i ⏐ 1 ≤ i ≤ κ ❵$ with $(i ≠ j ⇒ V_i ≠ V_j)$. We denote $\Rw = \{0, 1, …, 2κ\}$. For $m,n ∈ \Rw$ we define
  \begin{align}
    m⊕n & = (m+n) \bmod (2κ+1) \\
    m⊖n & = \min❨(m-n) \bmod (2κ+1), (n-m) \bmod (2κ+1)❩ \,.
  \end{align}
  We have $0 ≤ m⊖n = n⊖m ≤ κ$ for all $m,n ∈ \Rw$. Also, $(m⊕n)⊖m = n$ for arbitrary $m ∈ \Rw$ and $1 ≤ n ≤ κ$.
  We define an irreflexive relation $⋖$ on $\Rw$
  \[
    m ⋖ n ⇔ (n-m) \bmod (2κ+1) < (m-n) \bmod (2κ+1) \,,
  \]
  such that for all different $m,n ∈ \Rw$ either $m ⋖ n$, or $n ⋖ m$. We have $m ⋖ m⊕n$ for arbitrary $m ∈ \Rw$ and $1 ≤  n ≤ κ$.
  \item $κ ≥ ω$. Let $\RUlt = ❴V_α ⏐ 0 < α < κ ❵$ with $(α ≠ β ⇒ V_α ≠ V_β)$. We denote $\Rw = κ$. For $α,β ∈ \Rw$ we define $α⊕β = α+β$ \footnote{Here `$+$' denotes ordinal addition. If $β<α$, we denote by $α-β$ the unique ordinal $γ$ such that $β+γ=α$.} and
  \[
    α⊖β = \begin{cases}
            α-β & \text{if $β<α$},\\
            β-α & \text{otherwise}.
          \end{cases}
  \]
  Again, we have $µ⊖ν = ν⊖µ$ for all $µ,ν ∈ \Rw$, and $(µ⊕ν)⊖µ = ν$ for arbitrary $µ ∈ \Rw$ and $0 < ν < κ$.
  As in the previous case, we define a relation $⋖$ on $\Rw$, which in this case is just the usual strict total order:
  \[
    µ ⋖ ν ⇔ µ < ν \,.
  \]
  We have $µ ⋖ µ⊕ν$ for arbitrary $µ ∈ \Rw$ and $0 < ν < κ$.
\end{enumerate}
\end{definition}

The domain of the canonical model is $W = \BUlt × \Rw$.

For each $p ∈ \VS$ we define $v(p) = ❴ x∈W ⏐ [p] ∈ x₁ ❵$.

We choose a set $\BUltz ⊆ \BUlt²$ such that for each $(U₁,U₂) ∈ \BUlt²$ it contains exactly one $x ∈ ❴(U₁,U₂), (U₂,U₁)❵$.

For each pair $(U₁,U₂) ∈ \BUlt²$ we choose one $V ∈ \RUlt$, such that
\[
  (U₁,U₂)∈R⁰_V ⨇ (U₁ = U₂ ⇒ V = V⁻¹) \,,
\]
and denote it by $V_{U₁,U₂}$.

For each $V ∈ \RUlt$ we define a relation $R_V ⊆ W²$:
\begin{align}
 &R_V = \Biggl\{ (x,y)∈W² \Biggm| \biggl( x₂≠y₂ ⨇ (x₁,y₁) ∈ R⁰_{V_{x₂⊖y₂}} ∩ R⁰_{V_{x₂⊖y₂}⁻¹} \biggr.\Biggr.\\*
 &\eqindent\eqindent\eqindent\eqindent\Biggl.\biggl. ⨇ ❪\!❨x₂⋖y₂ ⨇ V=V_{x₂⊖y₂}❩ ⨈ ❨y₂⋖x₂ ⨇ V=V_{x₂⊖y₂}⁻¹❩\!❫\!\biggr) \Biggr.\\*
 &\eqindent  \Biggl. ⨈ \biggl(x₂≠y₂ ⨇ (x₁,y₁) ∈ R⁰_{V_{x₂⊖y₂}} ∖ R⁰_{V_{x₂⊖y₂}⁻¹} ⨇ V=V_{x₂⊖y₂}\biggr) \Biggr.\\*
 &\eqindent  \Biggl. ⨈ \biggl(x₂≠y₂ ⨇ (x₁,y₁) ∈ R⁰_{V_{x₂⊖y₂}⁻¹} ∖ R⁰_{V_{x₂⊖y₂}} ⨇ V=V_{x₂⊖y₂}⁻¹\biggr) \Biggr.\\*
 &\eqindent  \Biggl. ⨈ \Biggl(\!\biggl(x₂=y₂ ⨈ (x₁,y₁) ∉ R⁰_{V_{x₂⊖y₂}} ∪ R⁰_{V_{x₂⊖y₂}⁻¹}\biggr) \Biggr.\Biggr.\\*
 & \Biggl.\Biggl. ⨇ \biggl(\!❪\!(x₁,y₁) ∈ \BUltz ⨇ V=V_{x₁,y₁}❫ ⨈ ❪\!(y₁,x₁) ∈ \BUltz ⨇ V=V_{y₁,x₁}⁻¹❫\!\biggr)\!\Biggr)\!\Biggr\}.
\end{align}

\begin{lemma}\label{th-properties-of-can-model}
  \begin{enumerate}
    \item\label{th-properties-of-can-model--union-is-all} $⋃ ❴ R_V ⏐ V ∈ \RUlt ❵ = W²$.
    \item\label{th-properties-of-can-model--no-intersection} $V'≠V''$ implies $R_{V'} ∩ R_{V''} = ∅$.
    \item\label{th-properties-of-can-model--converse} $R_{V⁻¹} = ❨R_V❩⁻¹$;
    \item\label{th-properties-of-can-model--RV-to-RVzero} $(x,y)∈R_V$ implies $(x₁,y₁)∈R⁰_V$;
    \item\label{th-properties-of-can-model--RVzero-to-RV} If $(U₁,U₂)∈R⁰_{V_ν}$, then for each $µ∈\Rw$ it holds that
    \[
      ❨(U₁,µ), (U₂,µ⊕ν)❩ ∈ R_{V_ν} \,.
    \]
  \end{enumerate}
\end{lemma}

\begin{proof}
  The first four items may be easily verified by considering the four cases in the definition. We prove the last one.
  Let $(U₁,U₂)∈R⁰_{V_ν}$ and $µ∈\Rw$. Note that $µ ⋖ µ⊕ν$ and hence $µ ≠ µ⊕ν$. Consider the pair $❨(U₁,µ), (U₂,µ⊕ν)❩$. We have $(µ⊕ν)⊖µ = µ⊖(µ⊕ν) = ν$. There are two possibilities:
  \begin{itemize}
   \item $(U₁,U₂) ∈ R⁰_{V_ν} ∩ R⁰_{V_ν⁻¹}$. As $µ ⋖ µ⊕ν$, we have $❨(U₁,µ), (U₂,µ⊕ν)❩ ∈ R_{V_ν}$.
   \item $(U₁,U₂) ∈ R⁰_{V_ν} ∖ R⁰_{V_ν⁻¹}$. Then $❨(U₁,µ), (U₂,µ⊕ν)❩ ∈ R_{V_ν}$.
  \end{itemize}
\end{proof}

For each $α∈\VR$ we define $R(α) = ⋃ ❴ R_V ⏐ V ∈ \RUlt ⨇ [α] ∈ V ❵$.

\begin{lemma}
  For each term $α ∈ \TR(\VR)$
  \[
    R(α) = ⋃ ❴ R_V ⏐ V ∈ \RUlt ⨇ [α] ∈ V ❵ \,.
  \]
\end{lemma}

\begin{proof}
  For each $(x,y) ∈ W²$ we denote by $V(x,y)$ the unique $V∈\RUlt$ such that $(x,y) ∈ R_V$. We need to prove that
  \[
    R(α) = ❴ (x,y) ∈ W² ⏐ [α] ∈ V(x,y) ❵ \,.
  \]
  This can be proved by structural induction on $α$.
\end{proof}

\begin{notation}
  For $a∈\TS(\VS)$ we denote $h(a) = ❴ U∈\BUlt ⏐ [a]∈U ❵$. \\
  For $α∈\TR(\VR)$ we denote $h(α) = ❴ V∈\RUlt ⏐ [α]∈V ❵$.
\end{notation}

\begin{lemma}\label{th-quantifier-change-in-maximal-theory}
  If $a,b∈\TS(\VS)$ and $α∈\TR(\VR)$, then:
  \[
    ∀∃(a,b)[α]∈S ⇔ ❨∀U∈h(a)❩ ❨∀[c]∈U❩ ❨∃∃(c,b)[α]∈S❩ \,.
  \]
\end{lemma}

\begin{proof}
  \rightproof Let $∀∃(a,b)[α]∈S$, $U∈h(a)$ and $[c]∈U$. Then $a∩c≠0 ∈ S$. By \axiomref{eq-axiom-ee-ae}, $∃∃(c,b)[α]∈S$.

  \leftproof Assume that $∀∃(a,b)[α]∉S$. Since $S$ is a rich theory, there is a set variable $p$ such that $a∩p=0∨∃∃(p,b)[α] ∉ S$. Hence $a∩p=0 ∉ S$ and $∃∃(p,b)[α] ∉ S$. Then $[a]∩[p]≠[0]$ and there is an ultrafilter $U ⊇ ❴[a],[p]❵$.
\end{proof}

\begin{lemma}
  For each formula $φ ∈ \Formulas(\VS,\VR)$ the following equivalence holds: $φ ∈ S ⇔ M ⊨ φ$.
\end{lemma}

\begin{proof}
  The proof is by induction on the structure of $φ$. Since $S$ is a maximal theory, we need to consider explicitly only the cases where $φ$ is an atomic formula.
  \begin{enumerate}
    \item $φ$ is $a₁≤a₂$. By the Stone representation theorem for Boolean algebras,
      \begin{multline}
        a₁≤a₂ ∈ S ⇔ h(a₁) ⊆ h(a₂) \\*
        {} ⇔ v(a₁) = h(a₁)×\Rw ⊆ h(a₂)×\Rw = v(a₂) \,.
      \end{multline}
    \item $φ$ is $∃∃(a₁,a₂)[α]$.

      \rightproof Let $∃∃(a₁,a₂)[α] ∈ S$. Then $❨[a₁), [a₂)❩ ∈ R⁰_{[α)}$. By \thref{th-filter-to-ultrafilter-extension}
      \[
        ❨∃U₁ ∈ h(a₁)❩ ❨∃U₂ ∈ h(a₂)❩ ❨∃V ∈ h(α)❩ ❨(U₁,U₂) ∈ R⁰_V❩ \,.
      \]
      Let $V = V_µ$. Then $(U₁,0) ∈ v(a₁)$, $(U₂,µ) ∈ v(a₂)$ and $R_V⊆R(α)$. By \itemref{th-properties-of-can-model--RVzero-to-RV} in \thref{th-properties-of-can-model}, $❨(U₁,0),(U₂,µ)❩ ∈ R_V$. This shows that $M ⊨ ∃∃(a₁,a₂)[α]$.

      \leftproof Let $M ⊨ ∃∃(a₁,a₂)[α]$. Then
      \[
        ❨∃x ∈ v(a₁)❩ ❨∃y ∈ v(a₂)❩ ❨∃V ∈ h(α)❩ ❨(x,y) ∈ R_V❩ \,.
      \]
      Thus, we have $[a₁] ∈ x₁$, $[a₂] ∈ y₁$, $[α] ∈ V$, and \itemref{th-properties-of-can-model--RV-to-RVzero} in \thref{th-properties-of-can-model} gives us $(x₁,y₁) ∈ R⁰_V$. Hence $∃∃(a₁,a₂)[α] ∈ S$.

    \item $φ$ is $∀∀(a₁,a₂)[α]$.
      \begin{align}
        & ∀∀(a₁,a₂)[α] ∈ S ⇔ ¬∃∃(a₁,a₂)[-α] ∈ S \\
        & \eqindent {} ⇔ ∃∃(a₁,a₂)[-α] ∉ S ⇔ M ⊭ ∃∃(a₁,a₂)[-α] \\
        & \eqindent {} ⇔ M ⊨ ∀∀(a₁,a₂)[α]
      \end{align}

    \item $φ$ is $∀∃(a₁,a₂)[α]$.

      \rightproof Let $∀∃(a₁,a₂)[α] ∈ S$. By \thref{th-quantifier-change-in-maximal-theory}
      \[
        ❨∀U₁ ∈ h(a₁)❩ ❨∀[c] ∈ U₁❩ ❨∃∃(c,a₂)[α]∈S❩ \,.
      \]
      Hence $❨∀U₁ ∈ h(a₁)❩ ❪ ❨U₁,[a₂)❩ ∈ R⁰_{[α)} ❫$. By \thref{th-filter-to-ultrafilter-extension}
      \[
        ❨∀U₁ ∈ h(a₁)❩ ❨∃U₂ ∈ h(a₂)❩ ❨∃V ∈ h(α)❩ ❨(U₁,U₂) ∈ R⁰_V❩ \,.
      \]
      By \itemref{th-properties-of-can-model--RVzero-to-RV} in \thref{th-properties-of-can-model},
      \begin{multline}
        ❨∀U₁ ∈ h(a₁)❩ ❨∃U₂ ∈ h(a₂)❩ ❨∃ν ∈ \Rw❩ ❨∀µ ∈ \Rw❩   \\*
                                              ❪ ❨(U₁,µ), (U₂,µ⊕ν)❩ ∈ R_{V_ν} ⨇ V_ν ∈ h(α) ❫
      \end{multline}
      and hence
      \begin{multline}
        ❨∀U₁ ∈ h(a₁)❩ ❨∀µ ∈ \Rw❩ ❨∃U₂ ∈ h(a₂)❩ ❨∃ν ∈ \Rw❩   \\*
                                                              ❪ ❨(U₁,µ), (U₂,µ⊕ν)❩ ∈ R(α) ❫
      \end{multline}
      Therefore $❨∀x ∈ v(a₁)❩ ❨∃y ∈ v(a₂)❩ ❨(x,y) ∈ R(α)❩$.

      \leftproof Assume that $M ⊨ ∀∃(a₁,a₂)[α]$. Then
      \[
        ❨∀x ∈ v(a₁)❩ ❨∃y ∈ v(a₂)❩ ❨∃V ∈ h(α)❩ ❨(x,y) ∈ R_V❩ \,.
      \]
      By \itemref{th-properties-of-can-model--RV-to-RVzero} in \thref{th-properties-of-can-model},
      \[
        ❨∀x ∈ v(a₁)❩ ❨∃y ∈ v(a₂)❩ ❨∃V ∈ h(α)❩ ❨(x₁,y₁) ∈ R⁰_V❩ \,.
      \]
      As $\Rw≠∅$,
      \[
        ❨∀U₁ ∈ h(a₁)❩ ❨∃U₂ ∈ h(a₂)❩ ❨∃V ∈ h(α)❩ ❨(U₁,U₂) ∈ R⁰_V❩ \,.
      \]
      Hence
      \[
        ❨∀U₁ ∈ h(a₁)❩ ❨∀[c] ∈ U₁❩ ❨∃∃(c,a₂)[α] ∈ S❩ \,.
      \]
      Then, by \thref{th-quantifier-change-in-maximal-theory} we obtain $∀∃(a₁,a₂)[α] ∈ S$.

    \item $φ$ is $∃∀(a₁,a₂)[α]$.
      \begin{align}
        & ∃∀(a₁,a₂)[α] ∈ S ⇔ ¬∀∃(a₁,a₂)[-α] ∈ S \\
        & \eqindent {} ⇔ ∀∃(a₁,a₂)[-α] ∉ S ⇔ M ⊭ ∀∃(a₁,a₂)[-α] \\
        & \eqindent {} ⇔ M ⊨ ∃∀(a₁,a₂)[α]
      \end{align}
  \end{enumerate}
\end{proof}

\begin{theorem}[Completeness]
  If $Γ⊆\Formulas(\VS,\VR)$, then
  \[
    Γ \text{ is consistent} ⇔ Γ \text{ has a model} \,.
  \]
\end{theorem}

\begin{proof}
\rightproof Let $Γ$ be a consistent set of formulas. By \itemref{th-every-consistent-set-is-contained-in-a-maximal-theory} in \thref{th-every-consistent-set-is-contained-in} there is a maximal theory $S$ which contains $Γ$. $S$ has a model which is also a model of $Γ$.

\leftproof Let $M⊨Γ$ and let $Δ = ❴ φ ∈ \Formulas(\VS,\VR) ⏐ M⊨φ ❵$. By \thref{th-soundness} we have $\Thm(\VS, \VR) ⊆ Δ$. As \ruleref{eq-rule-MP} preserves the truth in every model, the set $Δ$ is closed under \ruleref{eq-rule-MP}. Since $M⊭⊥$, $⊥∉Δ$. Hence $Γ ⊆ Δ$ is consistent.
\end{proof}

\section{Complexity}\label{sect-complexity}

Before we consider the complexity of the satisfiability problem, we first note that our logic is a fragment of Boolean Modal Logic (BML)\cite{bml1,bml2} extended with a symbol `$⁻¹$' for the converse of the accessibility relation.

BML is a multimodal logic, whose language contains two types of variables -- a set of relational variables (atomic modal parameters) and an infinite set of propositional variables. The set of modal parameters consists of the set of relational variables, the relational constant $1$ and all their Boolean combinations. The set of formulas is the smallest set which contains the propositional variables and is closed under prefixing a formula by a box or diamond modality as well as connecting formulas with the propositional operators.

A model for BML is a triple $M = (W, R, v)$, where $W≠∅$ is the domain, $R$ is a function, which assigns to each atomic modal parameter a relation on $W$, and $v$ assigns to each propositional variable a subset of $W$. $R$ is extended to all modal parameters according to the standard interpretation of the Boolean operators as set intersection, union and complement, interpreting $1$ as the universal relation $W²$. We have the standard meaning of the modal operators:
\begin{align}
 & (M,x) ⊨ \modaldiamond{α}φ ⇔ (∃y∈W) ❨(x,y)∈R(α) ⨇ (M,y)⊨φ❩ \\
 & (M,x) ⊨ \modalbox{α}φ ⇔ (∀y∈W) ❨(x,y)∈R(α) ⇒ (M,y)⊨φ❩ \,.
\end{align}

We consider the extension of the language of BML with a symbol `$⁻¹$' in modal parameters. We interpret it as taking the converse of the relation: $R(α⁻¹)=❨R(α )❩⁻¹$. The formulas of our language have equivalents in this extension of the language of BML:

\begin{tabular}{r@{ is equivalent to }l}
  $a≤b$          & $\modalbox{1}(a→b)$      \\
  $∃∃(a,b)[α]$ & $\modaldiamond{1}❨a∧\modaldiamond{α}b❩$   \\
  $∀∃(a,b)[α]$ & $\modalbox{1}❨a→\modaldiamond{α}b❩$   \\
  $∀∀(a,b)[α]$ & $\modalbox{1}❨a→\modalbox{-α}¬b❩$ \\
  $∃∀(a,b)[α]$ & $\modaldiamond{1}❨a∧\modalbox{-α}¬b❩$.
\end{tabular}

The satisfiability problem for our logic is decidable in NExpTime, since the formulas are translatable (in polynomial time) into the NExpTime-decidable two-variable fragment of first-order predicate logic. We argue that the complexity is the same as the complexity of BML, which is proved by Lutz and Sattler \cite{ls} to be NExpTime if the language contains an infinite number of relational variables, and ExpTime if only a finite number of relational variables is available. Also, the complexity does not depend on whether we allow $⁻¹$ in the language.

In the case of an infinite number of relational variables, the lower NExpTime bound is proved in \cite{ls} by a reduction from an NExpTime-complete tiling problem. The BML formula, used to encode the tiling, is a conjunction of a formula, which describes the initial condition for the problem, and several conjuncts, which ensure that every model satisfying the formula is indeed a tiling. All conjuncts but the one for the initial condition can be translated into our fragment. The formula for the initial condition can be replaced by a formula from our fragment, such that the whole conjunction is equisatisfiable with the original one.

In the case of a finite number of relational variables, the lower ExpTime bound of BML follows from the ExpTime-completeness of $K_u$ (the basic modal logic enriched with the universal modality). However, the intersection of $K_u$ with our fragment is also ExpTime-hard, hence the ExpTime-hardness of our logic.

The upper ExpTime bound for BML is proved in \cite{ls} by reduction to the satisfiability problem for the basic multimodal logic enriched with the universal modality. The same reduction is applicable in the presence of $⁻¹$, and multimodal $K_u$ enriched with $⁻¹$ is also ExpTime-complete.

These high complexities are due to the presence of $∀∃$ and $∃∀$ in the language. If we remove these symbols from the language, the resulting logic has an NP-complete satisfiability problem, as it possesses the polysize model property. This can be proved by selection of points from a model in the way it is done in \cite{dlrbts} for the dynamic logics of the region-based theory of discrete spaces.

\section{Concluding remarks}\label{conclusions}

The first completeness proof for a non-classical relational syllogistic (i.e. one that contains relational terms) was given by Nishihara, Morita, and Iwata in \cite{NMI}. Their fragment contains variables for proper nouns and n-ary relational terms closed only under complementation and does not allow Boolean operations on set terms.

Later works on relational syllogistics, devoted mainly to the computational complexity problems, are McAllester and Givan \cite{Givan} and Pratt-Hartmann \cite{Pratt,Pratt1,Pratt2,Pratt-Third}. The paper by Moss and Pratt-Hartmann \cite{PrattMoss} is devoted both to complete axiomatizations and some computational complexity results. A successor of \cite{PrattMoss} is Moss \cite{Moss3}, devoted to axiomatizations and completeness proofs for a number of relational syllogistics.

Our logic differs in expressiveness from all systems of relational syllogistic mentioned above. One of the reasons is that we have quite rich language based on both class terms and relational terms, while the other logics are based on languages that are weaker than our system, or incomparable with it, some of them dealing only with atomic formulas. Such is, for instance, the system of McAllester and Givan \cite{Givan} and some systems studied in Moss and Pratt-Hartmann \cite{PrattMoss} and Moss \cite{Moss3}. The fragment of our language, which contains only two relational terms $α$ and $-α$ and all set terms are variables or negated variables, coincides with the language of the system $\mathcal{R}^{\dag}$ studied by Moss and Pratt-Hartmann in \cite{PrattMoss}.

In the present paper we have proved the completeness of a syllogistic logic with a set of binary relations closed under the Boolean operations and under taking the converse. The completeness proof can be generalized to the case of $n$-ary relations for arbitrary $n$, which will cover the case of $n$-transitive verbs. We also plan to study extensions of our logic with several kinds of nominals making it possible to cover sentences from natural language like `Socrates is a man', `Socrates is mortal'.

The construction of the canonical model in our logic is similar to that for BML. It is also possible to use the construction of the relations $R_V$ from $R⁰_V$ to prove the completeness of BML extended with a symbol $⁻¹$ for the converse of the accessibility relation.

\section*{Acknowledgements}

The authors would like to thank Ian Pratt-Hartmann and the anonymous referees for valuable comments on the paper.

This work was supported by the European Social Fund through the Human Resource Development Operational Programme 2007--2013 under contract BG051PO001-3.3.04/27/ 28.08.2009, by the project DID02/32/2009 of Bulgarian Science Fund and by Sofia University under contract 136/2010.

\appendix

\section{Proof of \thref{th-rel-bool-algebra}}\label{Appendix-proof-of-th-rel-bool-algebra}

In the proof of \thref{th-rel-bool-algebra} we will need the following lemmas.

If $Q$ is a quantifier, we will denote by $\overline{Q}$ the dual quantifier.

\begin{lemma}\label{th-dual-quantifiers}
  Let $a$ and $b$ be set terms and let $α$ be a relational term. Then, for arbitrary quantifiers $Q₁$ and $Q₂$ the formula
  \[
     Q₁\,Q₂ (a,b)[-α] ↔ ¬ \overline{Q₁}\,\overline{Q₂} (a,b)[α]
  \]
  is a theorem.
\end{lemma}

\begin{proof}
  One of these four formulas is an axiom. It remains to prove 6 implications. In the following proofs $p$ and $q$ are different set variables, which do not occur in the terms $a$ and $b$. We know that such variables exist, since $\VS$ is infinite.
  \begin{enumerate}
    \item\label{th-dual-quantifiers-item-ae-not-ea-minus} $\begin{aligned}[t]
      ⊢ & ∀∃(a,b)[α] → a∩p=0 ∨  ∃∃(p,b)[ α]          && \byaxiom{eq-axiom-ee-ae} \\
      ⊢ & ∀∃(a,b)[α] → a∩p=0 ∨ ¬∀∀(p,b)[-α]          && \byaxiom{eq-axiom-complement} \\
      ⊢ & ∀∃(a,b)[α] →         ¬∃∀(a,b)[-α]          && \byrule{eq-rule-not-aa-not-ea}
    \end{aligned}$
    \item\label{th-dual-quantifiers-item-not-ea-minus-ae} $\begin{aligned}[t]
      ⊢ & ¬∃∀(a,b)[-α] → a∩p=0 ∨ ¬∀∀(p,b)[-α]        && \byaxiom{eq-axiom-not-aa-not-ea} \\
      ⊢ & ¬∃∀(a,b)[-α] → a∩p=0 ∨  ∃∃(p,b)[ α]        && \byaxiom{eq-axiom-complement} \\
      ⊢ & ¬∃∀(a,b)[-α] →          ∀∃(a,b)[ α]        && \byrule{eq-rule-ee-ae}
    \end{aligned}$
    \item\label{th-dual-quantifiers-item-not-ee-minus-aa} $\begin{aligned}[t]
      ⊢ & ¬∃∃(a,b)[-α] → a∩p=0 ∨         ¬∀∃(p,b)[-α] && \byaxiom{eq-axiom-ee-ae} \\
      ⊢ & ¬∃∃(a,b)[-α]                                && \\*
        & \eqindent {} → a∩p=0 ∨ b∩q=0 ∨ ¬∀∀(p,q)[-α] && \byaxiom{eq-axiom-ae-aa} \\
      ⊢ & ¬∃∃(a,b)[-α] → a∩p=0 ∨ b∩q=0 ∨  ∃∃(p,q)[ α] && \byaxiom{eq-axiom-complement} \\
      ⊢ & ¬∃∃(a,b)[-α] →         b∩q=0 ∨  ∀∃(a,q)[ α] && \byrule{eq-rule-ee-ae} \\
      ⊢ & ¬∃∃(a,b)[-α] →                  ∀∀(a,b)[ α] && \byrule{eq-rule-ae-aa}
    \end{aligned}$
    \item\label{th-dual-quantifiers-item-not-ea-ae-minus} $\begin{aligned}[t]
      ⊢ & ¬∃∀(a,b)[α] → a∩p=0 ∨ ¬∀∀(p,b)[ α]          && \byaxiom{eq-axiom-not-aa-not-ea} \\
      ⊢ & ¬∃∀(a,b)[α] → a∩p=0 ∨  ∃∃(p,b)[-α]          && \byitem{th-dual-quantifiers-item-not-ee-minus-aa} \\
      ⊢ & ¬∃∀(a,b)[α] →          ∀∃(a,b)[-α]          && \byrule{eq-rule-ee-ae}
    \end{aligned}$
    \item\label{th-dual-quantifiers-item-aa-not-ee-minus} $\begin{aligned}[t]
      ⊢ & ∀∀(a,b)[α] → b∩p=0 ∨  ∀∃(a,p)[  α]          && \byaxiom{eq-axiom-ae-aa} \\
      ⊢ & ∀∀(a,b)[α] → b∩p=0 ∨ ¬∃∀(a,p)[ -α]          && \byitem{th-dual-quantifiers-item-ae-not-ea-minus} \\
      ⊢ & ∀∀(a,b)[α] → b∩p=0 ∨  ∀∃(a,p)[--α]          && \byitem{th-dual-quantifiers-item-not-ea-ae-minus} \\
      ⊢ & ∀∀(a,b)[α] →          ∀∀(a,b)[--α]          && \byrule{eq-rule-ae-aa} \\
      ⊢ & ∀∀(a,b)[α] →         ¬∃∃(a,b)[ -α]          && \byaxiom{eq-axiom-complement}
    \end{aligned}$
    \item\label{th-dual-quantifiers-item-ae-minus-not-ea} $\begin{aligned}[t]
      ⊢ & ∀∃(a,b)[-α] → a∩p=0 ∨  ∃∃(p,b)[-α]          && \byaxiom{eq-axiom-ee-ae} \\
      ⊢ & ∀∃(a,b)[-α] → a∩p=0 ∨ ¬∀∀(p,b)[ α]          && \byitem{th-dual-quantifiers-item-aa-not-ee-minus} \\
      ⊢ & ∀∃(a,b)[-α] →         ¬∃∀(a,b)[ α]          && \byrule{eq-rule-not-aa-not-ea}
    \end{aligned}$
  \end{enumerate} ~
\end{proof}

\begin{lemma}\label{th-ee-aa}
  Let $α,β ∈ \TR(\VR)$ and let $B = \TS(\VS)$. Then the following conditions are equivalent:
  \begin{enumerate}
    \item\label{th-ee-aa-ee} $(∀a,b ∈ B) ❨∃∃(a,b)[α]→∃∃(a,b)[β] ∈ S❩$
    \item\label{th-ee-aa-ae} $(∀a,b ∈ B) ❨∀∃(a,b)[α]→∀∃(a,b)[β] ∈ S❩$
    \item\label{th-ee-aa-aa} $(∀a,b ∈ B) ❨∀∀(a,b)[α]→∀∀(a,b)[β] ∈ S❩$
    \item\label{th-ee-aa-ea} $(∀a,b ∈ B) ❨∃∀(a,b)[α]→∃∀(a,b)[β] ∈ S❩$
  \end{enumerate}
\end{lemma}

\begin{proof}
  We will prove (\ref{th-ee-aa-ee})$→$(\ref{th-ee-aa-ae}). Assume that \itemref{th-ee-aa-ee} is true and suppose that there are set terms $a$ and $b$ such that $∀∃(a,b)[α]→∀∃(a,b)[β] ∉ S$. Since $S$ is a rich theory, there is a set variable $p$ such that
  \[
    ∀∃(a,b)[α] → a∩p=0 ∨ ∃∃(p,b)[β] ∉ S \,.
  \]
  Hence $∀∃(a,b)[α] → a∩p=0 ∨ ∃∃(p,b)[α] ∉ S$.

  This is a contradiction, since the last formula is a theorem.

  (\ref{th-ee-aa-ae})$→$(\ref{th-ee-aa-aa}) The proof is analogous to the previous one.

  (\ref{th-ee-aa-aa})$→$(\ref{th-ee-aa-ea}) Assume that \itemref{th-ee-aa-aa} is true. By \thref{th-dual-quantifiers}
  \[
    (∀a,b ∈ B) ❨¬∃∃(a,b)[-α]→¬∃∃(a,b)[-β] ∈ S❩ \,.
  \]
  For the sake of contradiction suppose that there are set terms $a$ and $b$ such that
  \[
    ∀∃(a,b)[-β]→∀∃(a,b)[-α] ∉ S \,.
  \]
  Since $S$ is a rich theory, there is a set variable $p$ such that
  \[
    ∀∃(a,b)[-β] → a∩p=0 ∨ ∃∃(p,b)[-α] ∉ S \,.
  \]
  Hence
  \[
    ∀∃(a,b)[-β] → a∩p=0 ∨ ∃∃(p,b)[-β] ∉ S \,.
  \]
  This is a contradiction, since the last formula is a theorem. Hence
  \[
    (∀a,b ∈ B) ❨∀∃(a,b)[-β]→∀∃(a,b)[-α] ∈ S❩ \,.
  \]
  Using \thref{th-dual-quantifiers} again, we conclude that
  \[
    (∀a,b ∈ B) ❨¬∃∀(a,b)[β]→¬∃∀(a,b)[α] ∈ S❩ \,,
  \]
  which implies \itemref{th-ee-aa-ea}.

  (\ref{th-ee-aa-ea})$→$(\ref{th-ee-aa-ee}) The proof is analogous to the previous one.
\end{proof}

\begin{lemma}\label{th-some-theorems}
  If $a$, $b$, $c$, $d$ are set terms and $α$, $β$ are relational terms, then the following formulas are theorems:
  \begin{enumerate}
    \item\label{th-some-theorems-ee-gt} $∃∃(a,b)[α] ∧ a≤c → ∃∃(c,b)[α]$ \\
      and \\
      $∃∃(a,b)[α] ∧ b≤c → ∃∃(a,c)[α]$
    \item\label{th-some-theorems-aa-lt} $∀∀(a,b)[α] ∧ c≤a → ∀∀(c,b)[α]$ \\
      and \\
      $∀∀(a,b)[α] ∧ c≤b → ∀∀(a,c)[α]$
    \item\label{th-some-theorems-intersection} $∀∀(a,b)[α] ∧ ∀∀(c,d)[β] → ∀∀(a∩c,b∩d)[α∩β]$
    \item\label{th-some-theorems-used-in-distibutive-axiom} $∀∀(a,b)[α] ∧ ¬∃∃(c,d)[α∩β] → ∀∀(a∩c,b∩d)[-β]$
  \end{enumerate}
\end{lemma}

\begin{proof} \leavevmode
  \begin{enumerate}
    \item We will prove the first one. \\
      $\begin{aligned}[t]
        ⊢ & ∃∃(a,b)[α] ∧ a≤c → ❨∃∃(a,b)[α] ∨ ∃∃(c,b)[α]❩ &&  \\
        ⊢ & ∃∃(a,b)[α] ∧ a≤c → ∃∃(a∪c,b)[α] ∧ a∪c=c      && \byaxiom{eq-axiom-union-in-first-argument-of-ee} \\
        ⊢ & ∃∃(a,b)[α] ∧ a≤c → ∃∃(c,b)[α]                && \byaxiom{eq-axiom-equality-first-argument}
      \end{aligned}$
    \item Follows from the previous item and \thref{th-dual-quantifiers}.
    \item $\begin{aligned}[t]
      ⊢ & ∀∀(a,b)[α] ∧ ∀∀(c,d)[β]                        && \\*
        & \eqindent {} → ∀∀(a∩c,b∩d)[α] ∧ ∀∀(a∩c,b∩d)[β] && \byitem{th-some-theorems-aa-lt}  \\
      ⊢ & ∀∀(a,b)[α] ∧ ∀∀(c,d)[β] → ∀∀(a∩c,b∩d)[α∩β]     && \byaxiom{eq-axiom-meet-conjunction}
    \end{aligned}$
    \item
    We will prove \itemref{th-some-theorems-used-in-distibutive-axiom}. Let $p$ and $q$ be different set variables which do not occur in $a$, $b$, $c$ and $d$. \\
    $\begin{aligned}[b]
      ⊢ & ∀∀(a,b)[α] ∧ ¬∃∃(c,d)[α∩β] ∧ ∀∀(p,q)[β]         && \\*
        & \eqindent {} → ∀∀(a∩p,b∩q)[α∩β] ∧ ¬∃∃(c,d)[α∩β] && \byitem{th-some-theorems-intersection}  \\
      ⊢ & ∀∀(a,b)[α] ∧ ¬∃∃(c,d)[α∩β] ∧ ∀∀(p,q)[β]         && \\*
        & \eqindent {} → a∩p∩c=0 ∨ b∩q∩d=0                && \byaxioms{eq-axiom-ee-ae}{eq-axiom-ae-aa}  \\
      ⊢ & ∀∀(a,b)[α] ∧ ¬∃∃(c,d)[α∩β]                      && \\*
        & \eqindent {} → a∩c∩p=0 ∨ b∩d∩q=0 ∨ ¬∀∀(p,q)[β]  &&   \\
      ⊢ & ∀∀(a,b)[α] ∧ ¬∃∃(c,d)[α∩β]                      && \\*
        & \eqindent {} → a∩c∩p=0 ∨ b∩d∩q=0 ∨ ∃∃(p,q)[-β]  && \byth{th-dual-quantifiers}  \\
      ⊢ & ∀∀(a,b)[α] ∧ ¬∃∃(c,d)[α∩β]                      && \\*
        & \eqindent {} → b∩d∩q=0 ∨ ∀∃(a∩c,q)[-β]          && \byrule{eq-rule-ee-ae}  \\
      ⊢ & ∀∀(a,b)[α] ∧ ¬∃∃(c,d)[α∩β]                      && \\*
        & \eqindent {} → ∀∀(a∩c,b∩d)[-β]                  && \byrule{eq-rule-ae-aa}
    \end{aligned}$ ~
  \end{enumerate}
\end{proof}

\mybeginproof{\thref{th-rel-bool-algebra}}
  The well-definition of $≤$ is obvious. The well-definition of $∪$ follows from \axiomref{eq-axiom-join-disjunction}. The well-definition of $∩$ and $-$ follows from \axiomref{eq-axiom-meet-conjunction}, \axiomref{eq-axiom-complement}, \thref{th-dual-quantifiers} and \thref{th-ee-aa}. The well-definition of $⁻¹$ follows from \axiomref{eq-axiom-converse}.

  We need to verify the following properties for arbitrary relational terms $α$, $β$ and $γ$:
  \begin{simplelist}
    \item\label{item-boolean-algebra-1} $α≼α$,\quad $(α≼β ⨇ β≼γ) ⇒ α≼γ$,\quad $(α≼β ⨇ β≼α) ⇒ α≈β$
    \item\label{item-boolean-algebra-2} $α∩β≼α$,\quad $α∩β≼β$,\quad $(γ≼α ⨇ γ≼β) ⇒ γ≼α∩β$
    \item\label{item-boolean-algebra-3} $α≼α∪β$,\quad $β≼α∪β$,\quad $(α≼γ ⨇ β≼γ) ⇒ α∪β≼γ$
    \item\label{item-boolean-algebra-4} $0_R≼α$,\quad $α≼1_R$
    \item\label{item-boolean-algebra-5} $α∩(β∪γ) ≼ (α∩β)∪(α∩γ)$
    \item\label{item-boolean-algebra-6} $α∩-α ≼ 0_R$
    \item\label{item-boolean-algebra-7} $1_R ≼ α∪-α$
  \end{simplelist}
  (\ref{item-boolean-algebra-1}) follows directly from the definition of the relation $≼$. (\ref{item-boolean-algebra-2}) follows from \axiomref{eq-axiom-meet-conjunction}. (\ref{item-boolean-algebra-3}) follows analogously from \axiomref{eq-axiom-join-disjunction}. (\ref{item-boolean-algebra-4}) follows from \axiomsref{eq-axiom-0_R}{eq-axiom-1_R}. We will prove the remaining three theorems. Let $a,b ∈ \TS(\VS)$.
  \begin{enumerate}
    \item[(\ref{item-boolean-algebra-5})] We will make use of \itemref{th-some-theorems-used-in-distibutive-axiom} in \thref{th-some-theorems}. Let $p,q ∈ \VS$, $p≠q$ and $\{p,q\} ∩ \Vs(a,b)=∅$.
    \begin{align}
      ⊢ & ∀∀(a,b)⟦α∩(β∪γ)⟧ ∧ ¬∃∃(p,q)⟦(α∩β)∪(α∩γ)⟧                        &&  \\*
        & \eqindent {} → ∀∀(a,b)[α] ∧ ∀∀(a,b)[β∪γ] ∧ ¬∃∃(p,q)[α∩β]        &&  \\*
        & \eqindent\eqindent {} ∧ ¬∃∃(p,q)[α∩γ]                           && \byaxioms{eq-axiom-meet-conjunction}{eq-axiom-join-disjunction} \\
      ⊢ & ∀∀(a,b)⟦α∩(β∪γ)⟧ ∧ ¬∃∃(p,q)⟦(α∩β)∪(α∩γ)⟧                        &&  \\*
        & \eqindent {} → ∀∀(a,b)[β∪γ] ∧ ∀∀(a∩p,b∩q)[-β] &&  \\*
        & \eqindent\eqindent {} ∧ ∀∀(a∩p,b∩q)[-γ] && \byth{th-some-theorems} \\
      ⊢ & ∀∀(a,b)⟦α∩(β∪γ)⟧ ∧ ¬∃∃(p,q)⟦(α∩β)∪(α∩γ)⟧                        &&  \\*
        & \eqindent {} → ∀∀(a,b)[β∪γ] ∧ ¬∃∃(a∩p,b∩q)[β] &&  \\*
        & \eqindent\eqindent {} ∧ ¬∃∃(a∩p,b∩q)[γ] && \byaxiom{eq-axiom-complement} \\
      ⊢ & ∀∀(a,b)⟦α∩(β∪γ)⟧ ∧ ¬∃∃(p,q)⟦(α∩β)∪(α∩γ)⟧                        &&  \\*
        & \eqindent {} → ∀∀(a,b)[β∪γ] ∧ ¬∃∃(a∩p,b∩q)[β∪γ]                 && \byaxiom{eq-axiom-join-disjunction} \\
      ⊢ & ∀∀(a,b)⟦α∩(β∪γ)⟧ ∧ ¬∃∃(p,q)⟦(α∩β)∪(α∩γ)⟧                        &&  \\*
        & \eqindent {} → a∩p=0 ∨ b∩q=0                                    && \byaxioms{eq-axiom-ee-ae}{eq-axiom-ae-aa} \\
      ⊢ & ∀∀(a,b)⟦α∩(β∪γ)⟧                                                &&  \\*
        & \eqindent {} → a∩p=0 ∨ b∩q=0 &&  \\*
        & \eqindent\eqindent {}  ∨ ∃∃(p,q)⟦(α∩β)∪(α∩γ)⟧             &&  \\
      ⊢ & ∀∀(a,b)⟦α∩(β∪γ)⟧                                                &&  \\*
        & \eqindent {} → b∩q=0 ∨ ∀∃(a,q)⟦(α∩β)∪(α∩γ)⟧                     && \byrule{eq-rule-ee-ae} \\
      ⊢ & ∀∀(a,b)⟦α∩(β∪γ)⟧ → ∀∀(a,b)⟦(α∩β)∪(α∩γ)⟧                         && \byrule{eq-rule-ae-aa}
    \end{align}
    \item[(\ref{item-boolean-algebra-6})] $\begin{aligned}[t]
      ⊢ & ∀∀(a,b)[α∩-α]                            && \\*
        & \eqindent {} → ∀∀(a,b)[α] ∧  ∀∀(a,b)[-α] && \byaxiom{eq-axiom-meet-conjunction} \\
      ⊢ & ∀∀(a,b)[α∩-α]                            && \\*
        & \eqindent {} → ∀∀(a,b)[α] ∧ ¬∃∃(a,b)[ α] && \byaxiom{eq-axiom-complement} \\
      ⊢ & ∀∀(a,b)[α∩-α] → a=0 ∨ b=0                && \byaxioms{eq-axiom-ee-ae}{eq-axiom-ae-aa} \\
      ⊢ & ∀∀(a,b)[α∩-α] → ¬∃∃(a,b)[-0_R]           && \byaxiom{eq-axiom-if-both-arguments-are-empty-then-not-ee} \\
      ⊢ & ∀∀(a,b)[α∩-α] →  ∀∀(a,b)[ 0_R]           && \byth{th-dual-quantifiers}
    \end{aligned}$
    \item[(\ref{item-boolean-algebra-7})] $\begin{aligned}[t]
      ⊢ & ¬∃∃(a,b)[α∪-α]                            && \\*
        & \eqindent {} → ¬∃∃(a,b)[α] ∧ ¬∃∃(a,b)[-α] && \byaxiom{eq-axiom-join-disjunction} \\
      ⊢ & ¬∃∃(a,b)[α∪-α]                            && \\*
        & \eqindent {} → ¬∃∃(a,b)[α] ∧  ∀∀(a,b)[ α] && \byth{th-dual-quantifiers} \\
      ⊢ & ¬∃∃(a,b)[α∪-α] → a=0 ∨ b=0                && \byaxioms{eq-axiom-ee-ae}{eq-axiom-ae-aa} \\
      ⊢ & ¬∃∃(a,b)[α∪-α] → ¬∃∃(a,b)[ 1_R]           && \byaxiom{eq-axiom-if-both-arguments-are-empty-then-not-ee} \\
      ⊢ &  ∃∃(a,b)[ 1_R] →  ∃∃(a,b)[α∪-α]           &&
    \end{aligned}$
  \end{enumerate} ~
  The equivalence $α ≼ β ⇔ α∪β ≈ β$ follows from \axiomref{eq-axiom-join-disjunction}.
\myendproof

\section{Proof of \thref{th-rel-ult-extension}}\label{Appendix-proof-of-th-rel-ult-extension}

  \begin{enumerate}
    \item This is obvious and is used only to shorten the notation.
    \item Let
    \begin{align}
      I & = \Bigl\{ [α] ∈ \RCl \Bigm| α ≈ α⁻¹ ⨇ ❨∃[a] ∈ U❩ ❨∃∃(a,a)[α] ∉ S❩ \Bigr\} \\
      F & = \Bigl\{ [α] ∈ \RCl \Bigm| α ≈ α⁻¹ \text{ and there exists a non-empty} \Bigr. \\*
      & \eqindent\eqindent \Bigl. \text{finite set } \{α₁,α₂,…,αₖ\} ⊆ \TR, \text{ such that} \Bigr. \\*
      & \eqindent\eqindent \Bigl. (α₁⁻¹∪-α₁) ∩ ⋯ ∩ (αₖ⁻¹∪-αₖ) ≼ α \Bigr\}
    \end{align}
    The set $I$ is an ideal in the Boolean algebra of symmetric classes of relational terms. By \thref{th-symmetry}, $F$ is a filter in that algebra and $F∩I = ∅$.

    By the separation theorem for filter-ideal pairs, there exists an ultrafilter $F' ⊇ F$ in the Boolean algebra of symmetric classes of relational terms, such that $F' ∩ I = ∅$. Let $V = ❴ x ∈ \RCl ⏐ x⁻¹∩x ∈ F' ❵$. $V$ has the following properties:
    \begin{itemize}
      \item $V ∈ \RUlt$. We need to check the following:
      \begin{enumerate}
	\item $[1_R] ∈ V$.
	\item If $x ∈ V$, $y ∈ \RCl$, and $x ≤ y$, then $y ∈ V$.
	\item If $x,y ∈ V$, then $x∩y ∈ V$.
	\item If $x ∈ \RCl$, then either $x ∈ V$, or $-x ∈ V$. Suppose otherwise. Then $x⁻¹∩x ∉ F'$ and $(-x⁻¹)∩-x ∉ F'$. Hence
	\[
	❨x⁻¹∩x❩ ∪ ❨(-x⁻¹)∩(-x)❩ = ❨x⁻¹∪-x❩ ∩ ❨(-x⁻¹)∪x❩ ∉ F' \,,
	\]
	which contradicts $F ⊆ F'$.
	\item $[0_R] ∉ V$.
      \end{enumerate}
      \item $V = V⁻¹$.
      \item $(U,U) ∈ R⁰_V$. Since $F' ∩ I = ∅$, we have
      \[
	❨∀[α] ∈ V❩ ❨∀[a] ∈ U❩ ❨∃∃(a,a)⟦α∩α⁻¹⟧ ∈ S❩
      \]
      and hence $(U,U) ∈ R⁰_V$.
    \end{itemize}
  \end{enumerate}

\end{document}